%% file: main.tex
\documentclass[graybox]{svmult}

\usepackage{type1cm}        
\usepackage{makeidx}         
\usepackage{graphicx}        
\usepackage{multicol}        
\usepackage[bottom]{footmisc}

\usepackage{hyperref}
\usepackage{newtxtext}       %
\usepackage[varvw]{newtxmath}       

\usepackage{acronym}
\acrodef{FEM}{Finite Element Method}
\acrodef{BREP}{Boundary REPresentation}
\acrodef{CAD}{Computer Aided Design}
\acrodef{NURBS}{Non-Uniform Rational B-Spline}
\acrodef{STEP}{STandard for the Exchange of Product model data}
\acrodef{ASP}{Answer Set Programming}
\acrodef{SJ}{Scaled Jacobian}
\acrodef{FGL}{Force directed Graph Layout}
\acrodef{PDE}{Partial Differential Equation}
\acrodef{NP}{Non-deterministic polynomial-time}

\definecolor{TUCblack}{rgb}{0,0,0} 
\definecolor{TUCwhite}{rgb}{1,1,1} 
\definecolor{TUCgreen}{rgb}{0,0.55,0.31} 
\definecolor{TUCgrey1}{rgb}{0.5,0.5,0.5} 
\definecolor{TUCgrey2}{rgb}{0.9,0.9,0.9} 
\definecolor{TUCred}{rgb}{0.55,0.11,0} 
\definecolor{TUCyellow}{rgb}{0.55,0.48,0} 
\definecolor{TUCdarkblue}{rgb}{0.1,0.26,0.35} 
\definecolor{TUCblue}{rgb}{0.2,0.48,0.65} 
\definecolor{TUCpurple}{rgb}{0.52,0.2,0.65} 

\makeindex             

\begin{document}

\title*{Combinatorial Methods in Grid based Meshing}
\author{Henrik Stromberg, Valentin Mayer-Eichberger and Armin Lohrengel}
\institute{Henrik Stromberg \at Institut für Maschinenwesen, Robert-Koch-Straße 32, 38678 Clausthal-Zellerfeld, Germany, \email{henrik@askdrq.com}
\and Valentin Mayer-Eichberger \at Name, Address of Institute \email{valentin@mayer-eichberger.de}
\and Armin Lohrengel \at Name, Address of Institute \email{lohrengel@imw.tu-clausthal.de}}
%
%
\maketitle

\abstract{This paper describes a novel method of generating hex-dominant meshes using pre-computed optimal subdivisions of the unit cube in a grid-based approach. 
Our method addresses geometries that are standard in mechanical engineering and often must comply with the restrictions of subtractive manufacturability. 
A central component of our method is the set of subdivisions we pre-compute with Answer Set Programming. Despite being computationally expensive, we obtain optimal meshes of up to 35 nodes available to our method in a template fashion.  
The first step in our grid-based method generates a coarse Precursor Mesh for meshing complete parts representing the bar stock. Then, the resulting mesh is generated in a subtractive manner by inserting and fitting the pre-generated subdivisions into the Precursor Mesh. 
This step guarantees that the elements are of good quality. In the final stage, the mesh nodes are mapped to geometric entities of the target geometry to get an exact match.
We demonstrate our method with multiple examples showing the strength of this approach.}

\section{Introduction}
\label{chapter:introduction}
Meshes containing elements of high quality are of uttermost importance for mechanical engineering applications using \ac{FEM}.
Current industry solutions require huge amount of manual geometry preprocessing and mesher configuration for \ac{FEM} solvers to produce accurate solutions with reasonable computation resources.
Our work addresses this issue by proposing a grid-based meshing approach that produces good quality hex-dominant meshes that also allow the element types prisms, pyramids and tetrahedra. Our novel method uses multiple algorithmic techniques such as heuristic and combinatorial methods. This paper explains our method, highlights the central concepts and demonstrates our method with examples. 

One key step of our algorithm uses the decompositions of a coarser grid into subdivisions of the elements that are intersected by the target geometry. 
The combinatorial explosion of this step is overcome by pre-computing all possible subdivisions using \ac{ASP}. We are able to compute optimal solutions for all required cases. 

The paper is structured as follows. In Section \ref{chapter:definitions} we define our terminology used throughout the paper and discuss mesh and element quality. Then in Section \ref{chapter:related_work} we discuss how our work relates to the state-of-art meshing algorithms. In Section \ref{chapter:algorithm} we describe our core method and in Section \ref{chapter:super_element_generation} we explain the generation of subdivisions and present statistics of their characteristic.
Section \ref{chapter:super_element_assignement} and Section \ref{chapter:entity_mapping} go into detail of the main steps of the algorithm. More examples are analysed and evaluated in Section \ref{chapter:results_on_exmples}. Finally, we conclude and discuss future work in Section \ref{chapter:futurework}.

\section{Background}
\label{chapter:definitions}
We choose the following terminology throughout this paper. 
The terms distinguish between entities of the mesh, its dual graph and the constraining geometry: 

\begin{table}[!t]
\caption{Term definitions}
\label{tab:term_definitions}

\begin{tabular}{p{5.65cm}p{5.65cm}}
\hline\noalign{\smallskip}
Term & Definition\\
\noalign{\smallskip}\svhline\noalign{\smallskip}

     Node &  Mesh node\\
     Graph node & Node of a Graph\\
     Edge & Mesh edge\\
     Graph edge & edge of a graph\\
     Face & Element face in the mesh\\
     Element & Element in the mesh\\
     Corner & Geometry node\\
     Curve & Geometry edge (may be curved)\\
     NURBS & Geometry surface\\
     Body & Geometry volume enclosed by\\ & NURBS\\
     Precursor Mesh & Mesh from which the part \\ & can be made by subtraction\\
     Retrenched  Mesh & Mesh after cutting of \\ & all nodes out of the target geometry\\
     Super Element & Element with twice the intended\\ &edge length to be decomposed into\\ &smaller sub elements \\&later in the pipeline\\

\noalign{\smallskip}\hline\noalign{\smallskip}
\end{tabular}

\end{table}

Our approach computes hex-dominant meshes with the application of \ac{FEM}. For structural applications, hex-dominant meshes can yield more accurate simulation results as opposed to fully hexahedral meshes as their elements are of higher quality (see e.g. \cite{veyhl_mesh_2010}). Fully hexahedral meshes often exhibit low quality elements in regions with high stresses, which reduces the overall solution quality.

Our work addresses quality of mesh elements for \ac{FEM} and we discuss established quality metrics.
Finite Elements must be convex and inversion free to be usable in this context. This property can be proved by showing that the Scaled Jacobian of all elements is positive \cite{Shepherd07}. Besides this requirement a wide range of mesh quality metrics exist. Furthermore, generated meshes shall be conformal and must not exhibit hanging nodes or edges.

\section{Related Work} \label{chapter:related_work}

This paper relates to multiple fields in the meshing community. 

Several approaches are known for the generation of quad meshes in \ac{FEM}. While octree based mesh generators can directly create volume meshes, Delaunay triangulation based meshers or whisker weaving based algorithms first create a surface mesh and then a volume mesh starting from the surface. Whisker weaving algorithms directly create hexahedral meshes.\\
The hexahedral meshes created through octree approaches cannot be used for \ac{FEM}, if they contain hanging nodes and result in non-conformal meshes. They can be removed by using so called octant cutout templates, which decompose each hexahedral octant in multiple tetrahedrons, thereby eliminating all hanging nodes. This was an inspiration for our approach working on hybrid meshes. Their resulting mesh is tetrahedral \cite{Yerry84}.\\
There are also octree based mesh generators which use all hexahedral octants such as by Borden et. al. \cite{Borden02}. However, implementations, which are available for productive use such as Snappyhexmesh, do not include techniques like the ones shown by \cite{Borden02}.\\
Delaunay based algorithms first create a triangulation and then try to generate hexahedrons from there (e.g. \cite{Remacle13}). Whisker weaving algorithms first create a volume mesh, based on a quadrilateral surface mesh, without defining node locations, thus only generating the mesh topology. Then node locations are set and faulty elements are resolved. The algorithms work well on simple box-like geometries, but struggles with more complex parts \cite{Tautges96}.

A field of active research is in algorithms that generate structured meshes. There are no unique definitions of structured meshes across the literature. In the context of our work, we use the definition of \cite{pietroni_hex-mesh_2022} and evaluate the structure of a mesh by examining its singularity graph or generally the valence of mesh nodes.
Thus, instead of a precise definition, the topic can be conquered with traits expected from a structured mesh. These traits are: low maximum valency (number of edges at a given node) over the whole mesh, a grid like structure and conservation of part symmetries.\\
The common way of acquiring structured meshes are sweeping algorithms. They create a volume mesh by sweeping a planar mesh through the volume (see e.g. \cite{Benzley99}). Sweeping mesh generators yield meshes with very good structural properties but are very limited in terms of accepted geometry. A common use case is to have the user dissect the geometry into sweepable sections. However, sweepability of the complete geometry is not guaranteed by all sections being sweepable and may also depend on the meshing order of the sections. This approach is widely used in commercial software such as ANSYS \cite{ansys}. It requires experienced users and often causes unexpected results.\\
A current development are quasi-structured quadrilateral meshing algorithms such as \cite{Reberol21} which relax the stated requirements for structured meshes and then can achieve an automated tool chain.

There are publications describing subtractive operations on meshes such as \cite{Borden02} or \cite{zhu13}. They perform a cutting operation on a given mesh by first altering node positions to have a closed loop of edges on the intersection curve between the initial mesh and the cutting tool. Then they remove elements and try to improve the resulting mesh.

Even though this approach was a starting point for the work presented here, it proved unusable to solve the problem of cutting a complex geometry out of a mesh. The main issue is that it does not guarantee to preserve part edges. A second issue is that the known method can create elements with critically low Scaled Jacobians. \cite{zhu13} shows an example where a chamfer is applied to a hex nut. The minimum Scaled Jacobian is low. Other publications such as \cite{Dhondt01} on subtractive meshing techniques actually perform a remeshing thus leaving the scope of the article.

Grid-based meshing algorithms such as the method described by Owen and Shelton in \cite{owen_evaluation_2015} use a grid of cubes which envelopes the desired geometry. Then boundary surfaces of the geometry are mapped onto the mesh. Finally, all sections of the mesh, which are outside of the target geometry, are cut from the mesh. Such methods yield good results on parts with high similarity of the generated grid to the target geometry but produce poor quality meshes on parts with deviating geometry. Their results are also sensitive to the orientation of the target geometry in global coordinates  \cite{pietroni_hex-mesh_2022}. 

Livesu et al. in \cite{Livesu21} proposes a similar grid-based method using Super Elements and refined subdivisions. However, their work considers only hexahedral elements. The computational problem to find subdivision becomes easier in their case and their method cannot guarantee minimal quality of the resulting elements. 

Many approaches with similar properties to the ones discussed exist, but follow the same general concepts. All known implementations of meshing algorithms use a \ac{BREP} geometry interface neglecting \ac{CAD} tree structure. Our method takes advantage of the expression from the \ac{CAD} tree and to the best of our knowledge this has not been before in the context of automatic hex-dominant meshing. 

Our study of subdivisions of the cube is related to other combinatorial approaches investigating similar problems:  
In \cite{Pellerin18} the computation of all possible decompositions of a single hexahedron into tetrahedrons is done through Boolean satisfiability solving, which use similar algorithmic techniques than ASP.  
Their method is limited to the 8 nodes of the hexahedron itself and introduces no further interior nodes.
In a related combinatorial approach, the decompositions of pyramids into hexahedra is studied in \cite{verhetsel_44-element_2019}. The use of Integer Programming in meshing is demonstrated in \cite{Pitzalis21} to optimize a balanced octree in a grid-based meshing algorithm.  

\section{Algorithm} \label{chapter:algorithm}

Our meshing algorithm consist of four stages: 

\begin{enumerate}
    \item Generate Precursor Mesh
    \item Assign Super Elements
    \item Map Mesh to Geometry
    \item Optimize Mesh
\end{enumerate}

A brief description of the steps is as follows: 
The first stage chooses a coarse Precursor Mesh from the target geometry as a starting point for the pipeline. Then, to each element in the Precursor Mesh a Super Element is assigned to roughly match the geometry restrictions. In the third step each node in the finer mesh of the Super Elements is mapped back to the target geometry. Finally, the mesh is optimized to improve its quality.   

To generate conformal meshes, the algorithm requires a complete collection of Super Element decompositions for any subset of the 8 corner nodes of the unit cube. This collection constitutes a main contribution of our work and the generation is explained in depth in Section \ref{chapter:super_element_generation}. In the rest of this section we discuss the prerequisites for the steps of our algorithm and our reasoning behind the proposed method.  

\begin{figure}
\begin{center}
\includegraphics[width=1.0\textwidth]{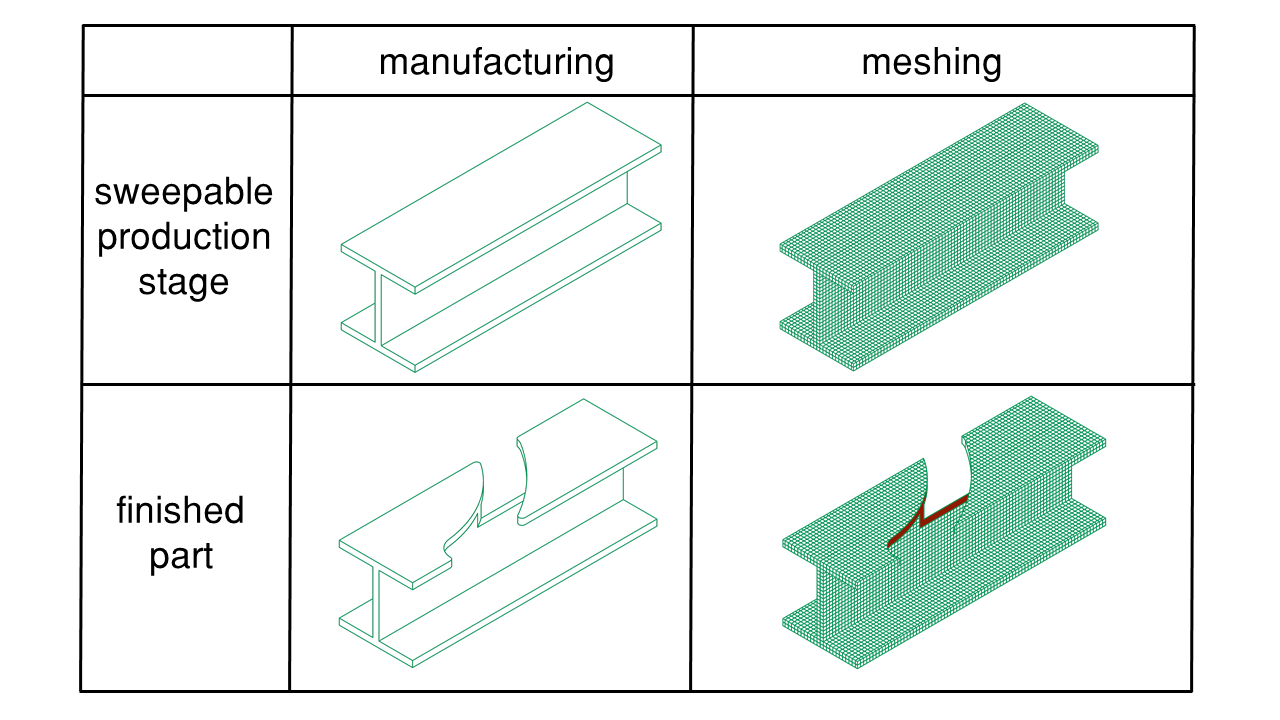}
\end{center}
\caption{Stages of part production typically for subtractive manufacturing in mechanical engineering.}
\label{fig:prod_stages}
\end{figure}

The costly computation for Super Elements does not scale to larger meshes with more than 35 nodes. Due to the size requirements of real-world applications of up to $10^8$ nodes, we have chosen an approach inspired by subtractive manufacturing. Many geometries meshed for simulations represent mechanical parts. Such parts are typically designed to be manufactured by removing material from bar stock such as a I-beam. As the bar stock material is a manufactured in a continuous rolling process, its geometry is easily meshable with sweeping methods. From this sweeped mesh the material has to be removed to carve the target geometry out of the bar stock, as shown in Figure \ref{fig:prod_stages}. The process stages are illustrated with a cube with a central cylindrical hole subtracted from it.

\begin{figure}
\sidecaption
\includegraphics[width=0.4\textwidth]{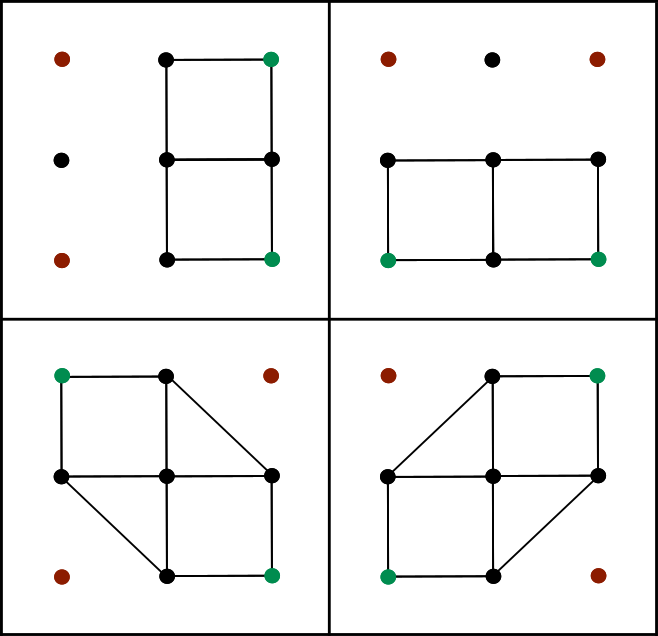}
\caption{Example of interfaces between Super Elements. Green nodes are retained Precursor Mesh nodes, dropped nodes are marked red. Possible additional nodes be they used or unused are drawn black.}
\label{fig:interfaces_between_superelements}
\end{figure}

\begin{figure}
\begin{center}
\includegraphics[width=0.5\textwidth]{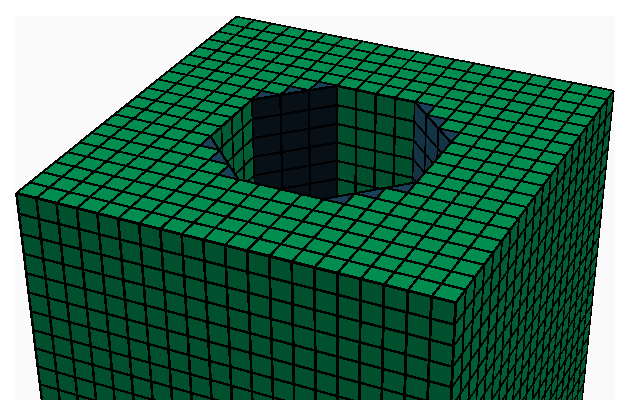}

\includegraphics[width=0.5\textwidth]{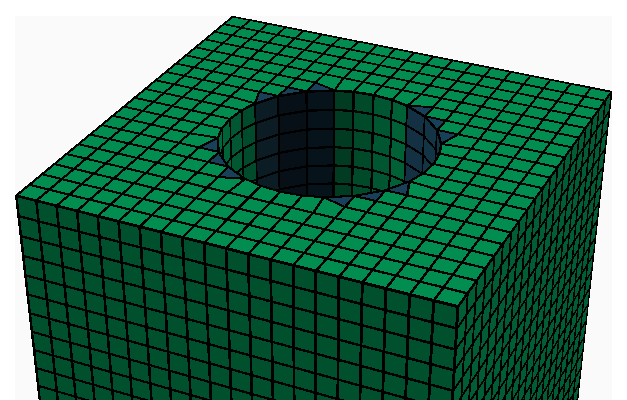}
\end{center}
\caption{Top: Mesh of Example 1 after assigning Super Elements. Bottom: Mesh after being mapped to geometry. This resulting mesh was automatically computed from target geometry with our proposed method and did not require manual improvements of the mesh. }
\label{fig:retrenched_mesh}
\end{figure}

Just removing elements from the mesh in order to emulate the cutting of material is unsufficient as doing so may create a rough surface which does not match the desired part contour. We have modeled the cutting of material by replacing all elements of the Precursor Mesh with pre-computed Super Elements.
All Super Elements we use are supposed to replace elements of the Precursor Mesh thereby increasing its granularity. Each Super Element is specified by removing some of the nodes from the base element geometry. For an eight node hexahedron $2^8=256$ possible Super Elements exist. In order to assure compatibility between the used Super Elements each node of the Precursor Mesh must be kept or removed for all adjacent elements and Super Elements for them must be chosen accordingly. Furthermore, we have globally defined the surface mesh of a Super Element face depending on the nodes present in the Super Element (see Figure \ref{fig:interfaces_between_superelements}). In this approach the assembly of a part from Super Elements can be seen as selecting which nodes of the Precursor Mesh are supposed to be present in the final mesh. We call the result of this process \emph{Retrenched  Mesh}. The Retrenched  Mesh for the used example is shown in Figure \ref{fig:retrenched_mesh} and the details of the process are described in Section \ref{chapter:super_element_assignement}. The node coordinates of the nodes within all Super Elements are transformed such that the outer nodes of the Super Element coincide with the corresponding nodes of the Precursor Mesh.

The resulting mesh is a rough approximation of the desired geometry. In order make the surfaces match, all faces of the surface mesh are bound to entities of the target geometry as shown for the example in Figure \ref{fig:retrenched_mesh}. The details of the process are described in Section \ref{chapter:entity_mapping}. Geometric entities, which are not fully represented, are deemed too small to be part of the mesh and are neglected as a consequence.


In a last step the mesh quality is improved by first optimizing the surface mesh and then the volume mesh. The surface mesh is optimized by moving surface nodes on their respective geometric entities. Nodes which are bound to corners cannot be moved. Currently elements with low \ac{SJ} are identified. Then the node locations of these elements are optimized one element at a time. The resulting changes are minuscule in the provided example.

The novel processes Super Element Assignment and Mesh Mapping can be implemented in $\mathcal{O}(n \log n)$ in the number of elements as shown in the respective sections. With this complexity of the complete algorithm we expect our approach to scale to typical industrial applications in simulations with up to $10^8$ nodes.

\section{Super Element Generation} \label{chapter:super_element_generation}

In this section we explain the application of Answer Set Programming to generate optimal Super Elements from subdivisions of the cube.

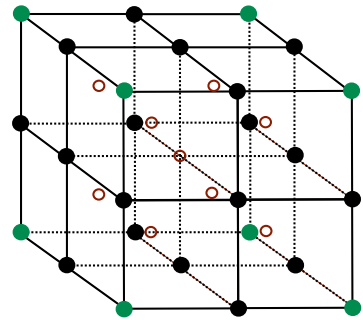
\begin{figure}
\sidecaption
\def\svgwidth{0.4\textwidth}
\input{pictures/internal_nodes_hex/internal_nodes_hex}
\caption{The set of all additional node positions for a hexahedron that are available for the generation of the subdivision. We consider all possible elements that can be generated from subsets of these nodes (Table \ref{tbl:element_portfolio}).}
\label{fig:hex_node_loc}
\end{figure}

To generate the more fine grained internal mesh of the hexahedral Super Elements, we chose additional 27 nodes located proportionally in the mid sections of the cube, in total being 35 node locations. The additional nodes provide a refinement of the Super Element. The node locations are shown in Figure \ref{fig:hex_node_loc}. The main task is to compute the optimal hybrid mesh that can be created from all these nodes. A naive brute-force algorithm would not be able to solve this computational task with realistic resources and more advanced methods are necessary.

With these node locations all possible tetrahedra, hexahedra, prisms and pyramids are generated and their Scaled Jacobian is computed according to \cite{Lobos15}. We have specified a minimum \ac{SJ} for each element type in order to exclude elements of poor quality and to speed up the solution progress. Ultimately, the element portfolio listed in table \ref{tbl:element_portfolio} is used to assemble mesh solutions. Note that the SJ values for different element types are incomparable and we selected different minimal thresholds. 

\begin{table}[!t]
\caption{Element portfolio}
\label{tbl:element_portfolio}

\begin{tabular}{p{3cm}p{3.4cm}p{2.5cm}p{2.4cm}}
\hline\noalign{\smallskip}
Element type & All elem. (SJ$>0$) & Min. \ac{SJ} & Count  \\
\noalign{\smallskip}\svhline\noalign{\smallskip}
Tetrahedrons & 44850 & 0.16 & 22750 \\ 
Hexahedrons & 16333575 & 0.3 & 3809 \\
Prisms & 1351685 & 0.45 & 2751 \\
Pyramids & 264501 & 0.35 & 1626 \\
\noalign{\smallskip}\hline\noalign{\smallskip}
\end{tabular}
\end{table}

We have modeled a logically consistent mesh in terms of its dual graph, enforcing the topological constraints with the following rule set:

\begin{itemize} 
    \item Each face of an element must connect to exactly one other element or be part of the outer hull.
    \item Any two neighboring faces on the outer hull are locally convex.  
    \item All elements are connected (there are no disconnected sub-graphs).
    \item No edge of any face may intersect any other face, except for shared nodes (the mesh is not self-intersecting).
\end{itemize}

All these constraints were modelled in \ac{ASP}, which is well suited for describing such graph-based constraints, and the program is natural and human readable.  
In particular, the connectivity constraint is difficult to formulate in related approaches such as Integer Programming or Boolean Satisfiability. 
We use the state-of-the-art ASP solver Clingo\footnote{https://potassco.org/clingo/} by the Potassco group \cite{Gebser19} to compute solutions to our problem.

There are 256 instances for which meshes have to be generated in order to get a complete set of Super Elements. Only 22 problems (plus the trivial empty mesh problem) have to be solved because many problems can be transformed into each other by rotating or mirroring the instance. Table \ref{tbl:element_portfolio} lists the number of elements with positive SJ for each type. Allowing all these elements in the optimization would generate too big instances. So, we carefully determined a minimal threshold for each element type to keep the problem instances manageable.    

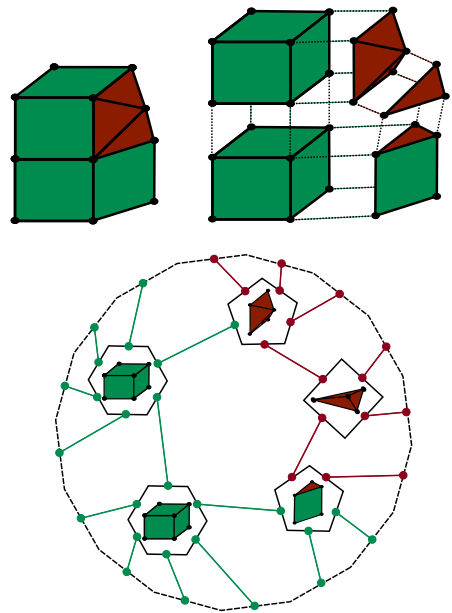
\begin{figure}
\sidecaption
\def\svgwidth{0.5\textwidth}
\input{pictures/mesh_graph/mesh_graph}
\caption{Graph representation of a mesh with triangular faces colored red and quadrangular faces colored green.}
\label{fig:graph_representation}
\end{figure}

Our implementation generates meshes for all cases in about one minute per problem with the element portfolio from Table \ref{tbl:element_portfolio} on 8 cores with 3.6 GHz. The solver proved optimality for all solutions.\\

\begin{table}[!t]
\caption{Optimization results for all subdivisions with best results for the three different criteria, respectively. Usually, the three resulting meshes per case are different. A comprehensive list of for the column Min SJ according to \cite{Lobos15} is shown in Figure \ref{tab:subdivisons}. SJs for mixed meshes cannot be compared to hexahedral meshes.}
\label{tbl:results}

\begin{tabular}{p{2.26cm}p{2.26cm}p{2.26cm}p{2.26cm}p{2.26cm}}
\hline\noalign{\smallskip}
        Case & \# Inst. & Max.     & Min SJ  & Max Elem. \\ 
             &           & Valence  &         & Count \\
\noalign{\smallskip}\svhline\noalign{\smallskip}
        1  & 8 & 3 & 0.35 & 1\\
        2  & 12 & 4 & 0.46 & 2\\
        3  & 12 & 8 & 0.26 & 20\\
        4  & 24 & 9 & 0.21 & 18\\
        5 & 6 & 5 & 1.00 & 4\\
        6 & 4 & 6 & 0.24 & 6\\
        7 & 24 & 13 & 0.21 & 36\\
        8 & 12 & 12 & 0.21 & 41\\
        9 & 8 & 8 & 0.29 & 14\\
        10 & 8 & 13 & 0.25 & 36\\
        11 & 12 & 13 & 0.21 & 41\\
        12 & 24 & 12 & 0.21 & 36\\
        13 & 24 & 9 & 0.21 & 28\\
        14 & 6 & 6 & 0.46 & 8\\
        15 & 24 & 12 & 0.35 & 29\\
        16 & 12 & 6 & 0.46 & 8\\
        17 & 2 & 22 & 0.35 & 52\\
        18 & 8 & 19 & 0.35 & 46\\
        19 & 12 & 16 & 0.35 & 40\\
        20 & 4 & 18 & 0.35 & 50\\
        21 & 8 & 12 & 0.35 & 29\\
        22 & 1 & 6 & 1.00 & 8\\
\noalign{\smallskip}\hline\noalign{\smallskip}
\end{tabular}
\end{table}

We are interested in various quality metrics of the Super Elements beyond maximizing the minimal SJ. To get a better idea if other criteria are better suited, we ran separate optimizations for the following goals:

\begin{itemize}
    \item maximize the minimum \ac{SJ} 
    \item minimize element count
    \item minimize the maximum node valency
\end{itemize}

\begin{figure*}
    \centering
    \begin{tabular}{cccc}
    \includegraphics[width=0.23\textwidth]{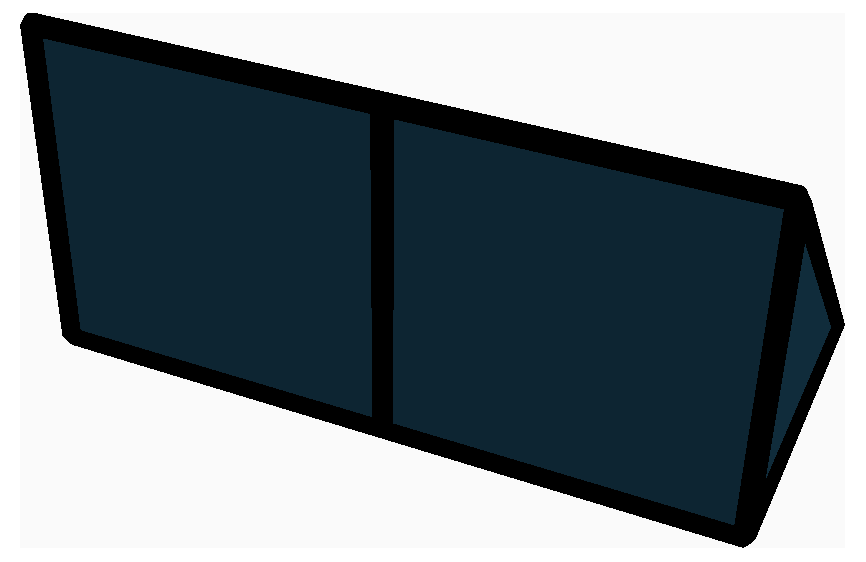}&
    \includegraphics[width=0.23\textwidth]{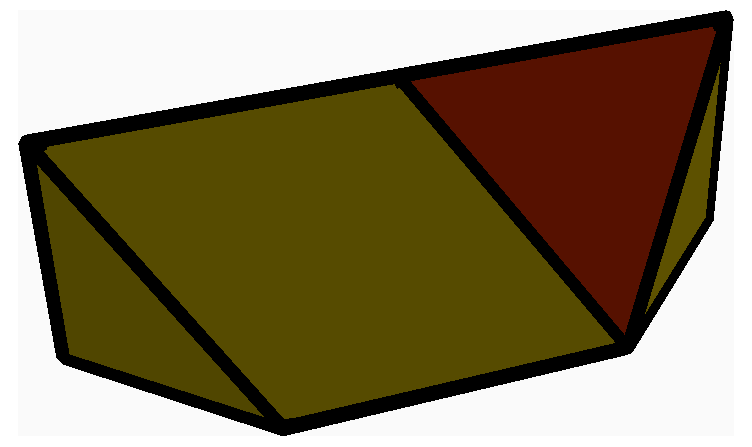}&
    \includegraphics[width=0.23\textwidth]{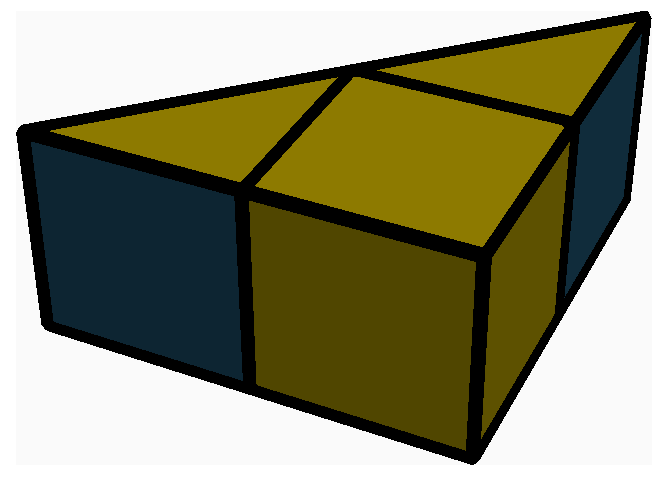}&
    \includegraphics[width=0.23\textwidth]{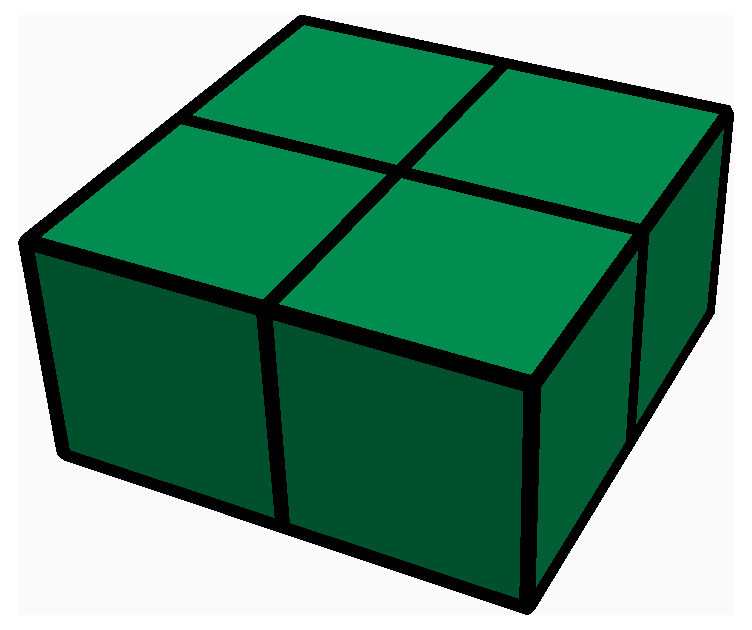}\\
    \includegraphics[width=0.23\textwidth]{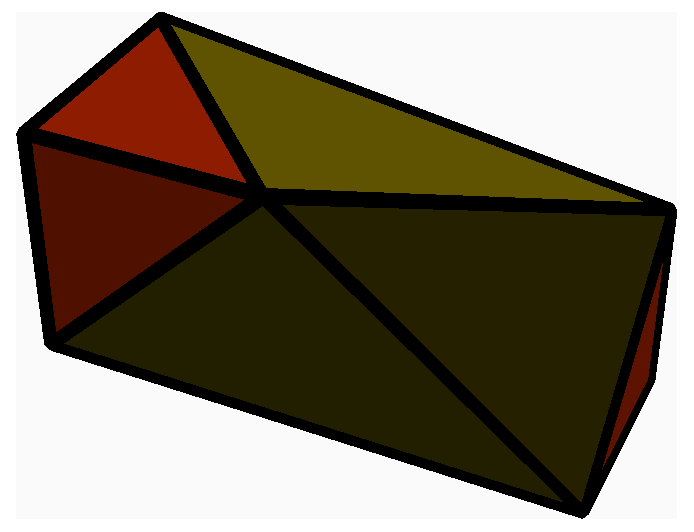}& 
    \includegraphics[width=0.23\textwidth]{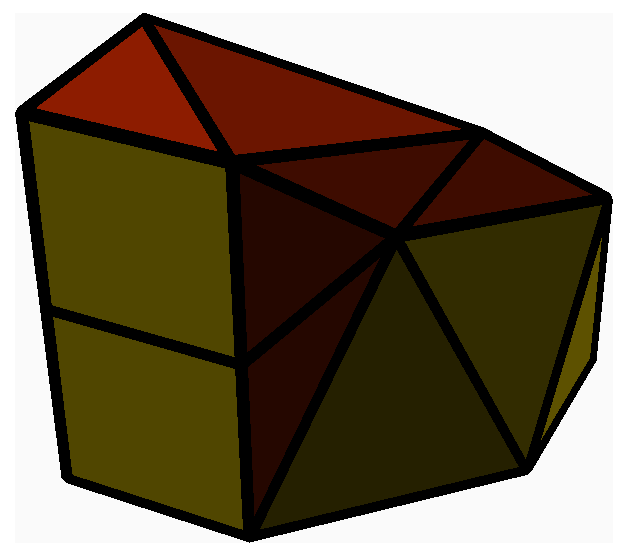}& 
    \includegraphics[width=0.23\textwidth]{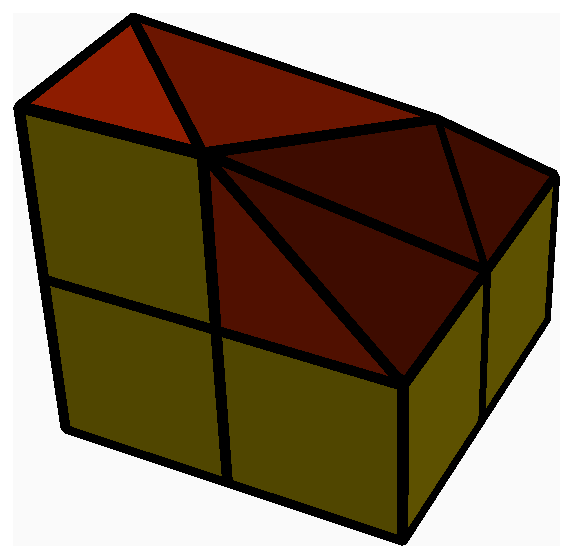}& 
    \includegraphics[width=0.23\textwidth]{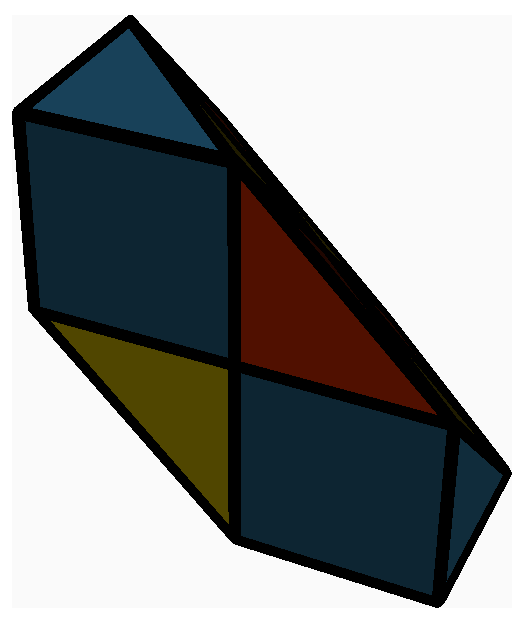}\\
    \includegraphics[width=0.23\textwidth]{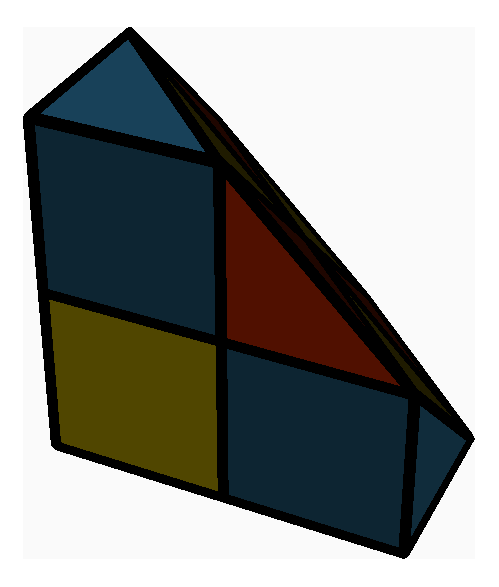}& 
    \includegraphics[width=0.23\textwidth]{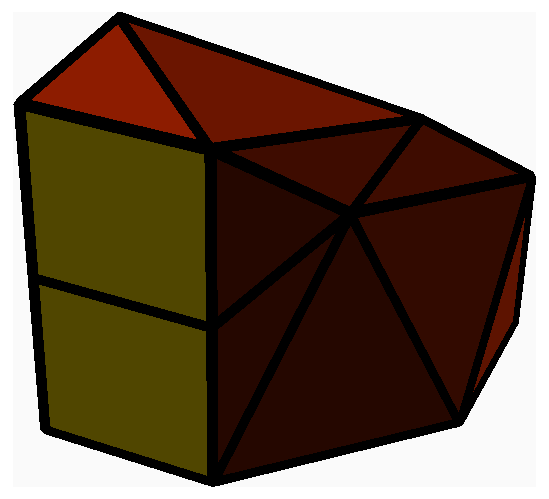}& 
    \includegraphics[width=0.23\textwidth]{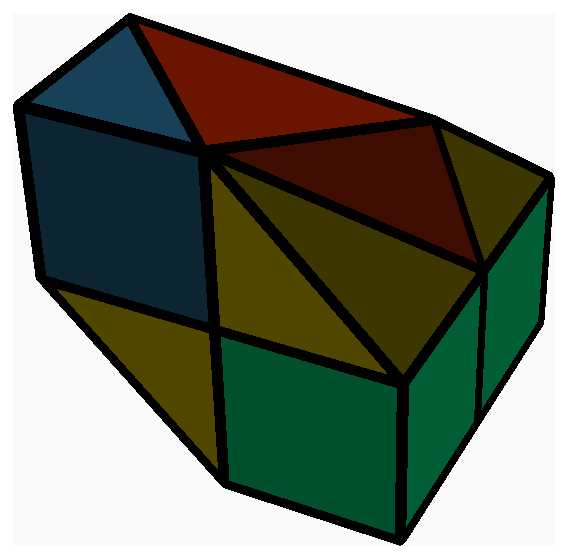}& 
    \includegraphics[width=0.23\textwidth]{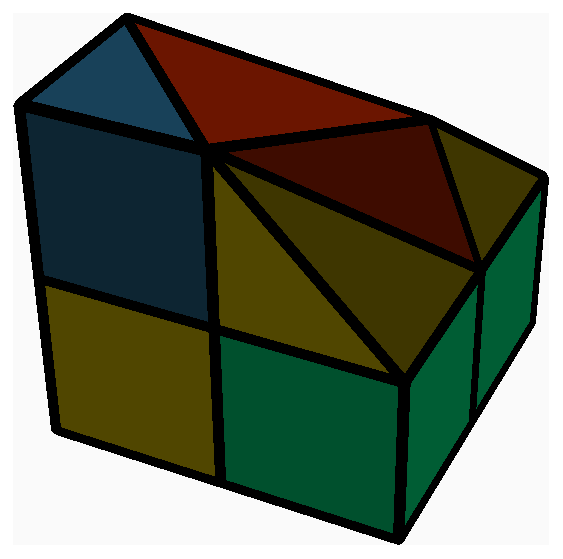}\\
    \includegraphics[width=0.23\textwidth]{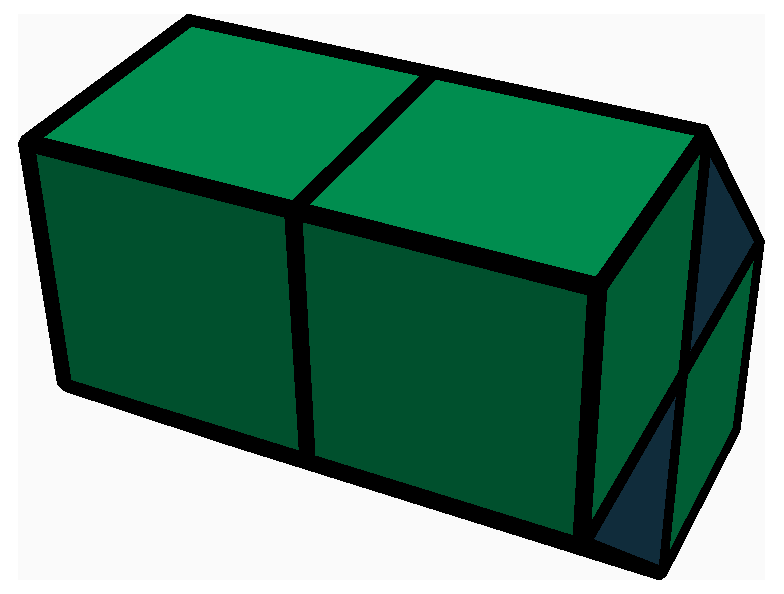}& 
    \includegraphics[width=0.23\textwidth]{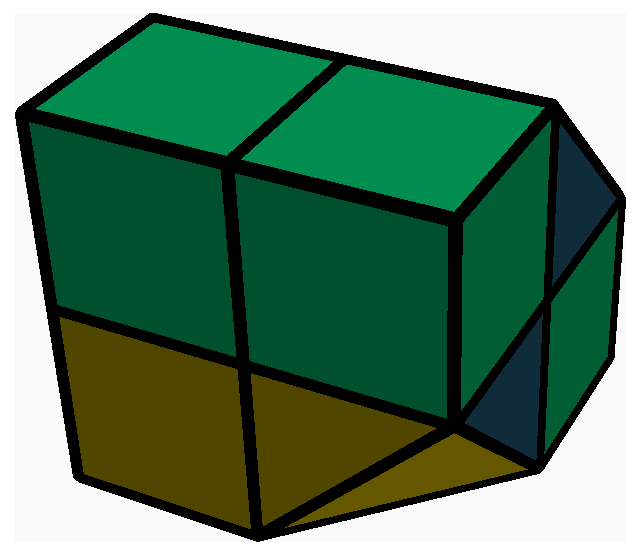}& 
    \includegraphics[width=0.23\textwidth]{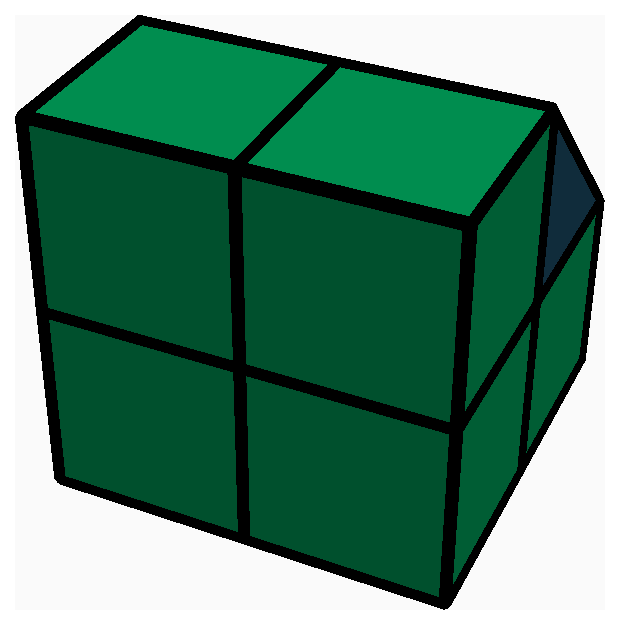}& 
    \includegraphics[width=0.23\textwidth]{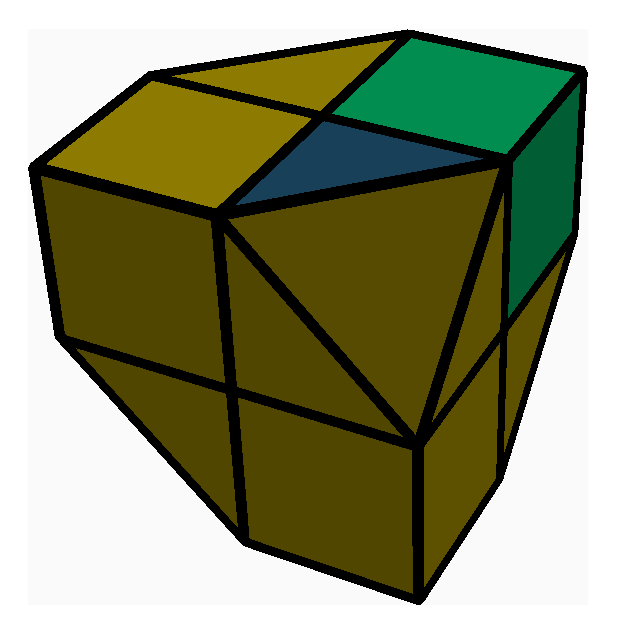}\\
    \includegraphics[width=0.23\textwidth]{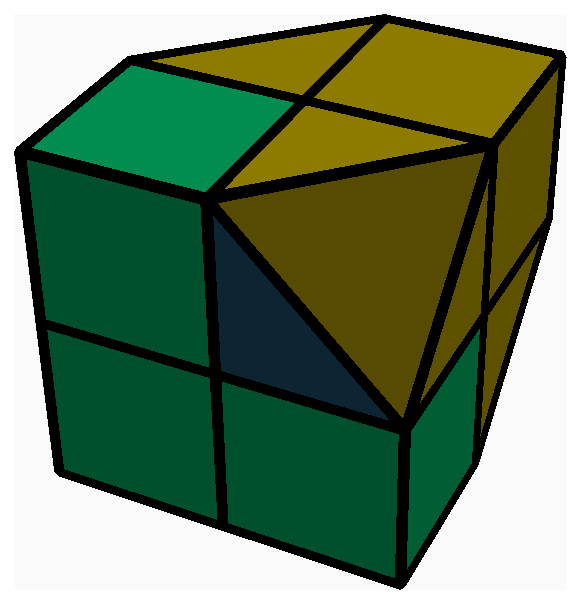}& 
    \includegraphics[width=0.23\textwidth]{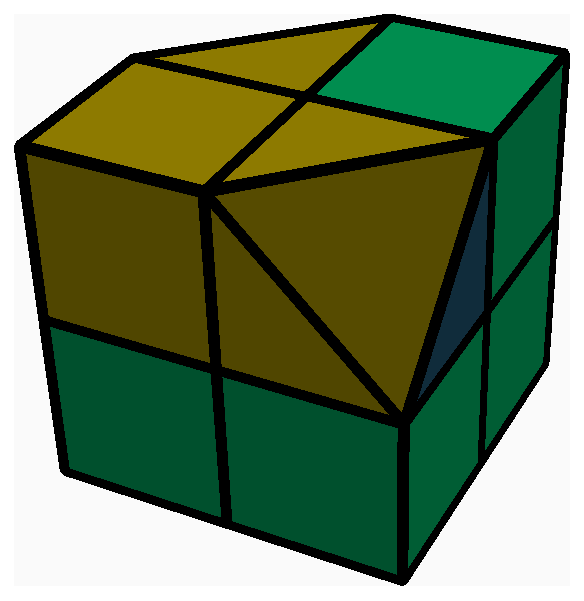}& 
    \includegraphics[width=0.23\textwidth]{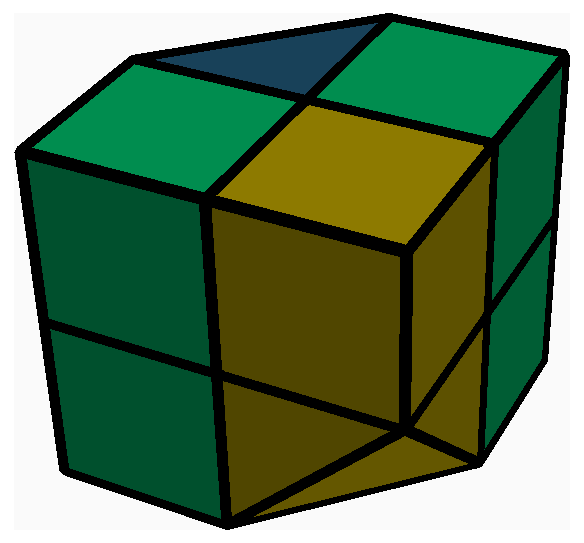}& 
    \includegraphics[width=0.23\textwidth]{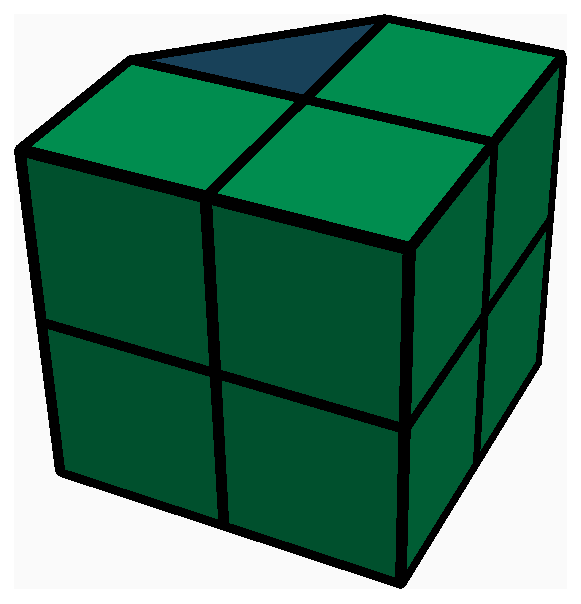}\\
    \end{tabular}
    \caption{All non-trivial subdivisions of a unit cube maximising the minimal SJ. The element types are indicated by the following color coding:  Hexahedra in {\color{TUCgreen}green},  Tetrahedra in {\color{TUCred}red}, Prisms in {\color{TUCblue}blue} and  Pyramids in {\color{TUCyellow}yellow}. }
    \label{tab:subdivisons}
\end{figure*}

For optimizing the minimum SJ we list all 20 non-trivial Super Elements in Figure \ref{tab:subdivisons}. For most of these solutions, it would be very hard to find these by hand, and impossible to prove that they are optimal. Table \ref{tbl:results} lists the optimization results for all 22 solved problems with values regarding the different goals. In the more complex instances, different goals lead to different meshes.  

\section{Super Element Assignment} \label{chapter:super_element_assignement}

For each element of the Precursor Mesh a Super Element has to be assigned. Compatibility between the selected Super Elements is guaranteed by deciding which nodes on the Precursor Mesh shall be present in the final mesh and selecting the resulting Super Element based on the present nodes. So the presence of each Precursor Mesh node in the final mesh can be described as a vector of Boolean variables $N$. 

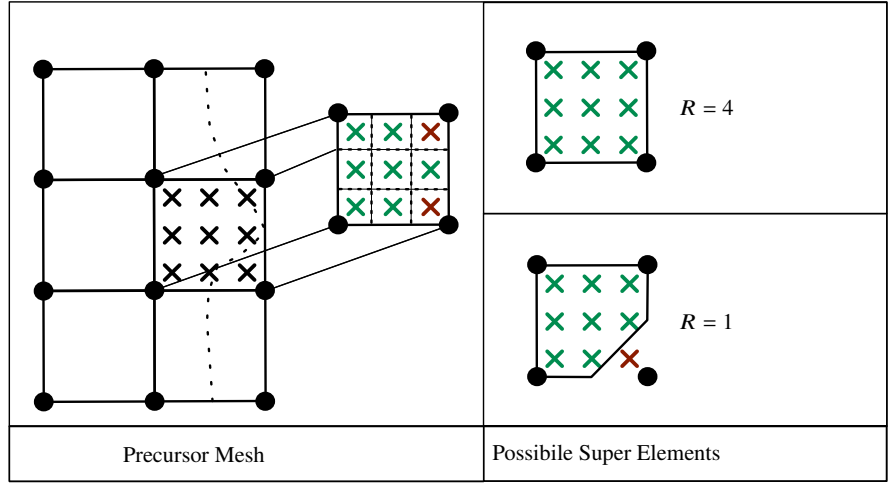
\begin{figure}
\centering
\def\svgwidth{1.0\textwidth}
\input{pictures/super_eleme_assignement/super_eleme_assignement}
\caption{Assembling a mesh by assigning Super Elements to Precursor Mesh elements}
\label{fig:super_element_assignment}
\end{figure}

For inserting each Super Element into each element of the Precursor Mesh a fit quality is computed by approximating the residual volume $R$. Fig. \ref{fig:super_element_assignment} illustrates this procedure simplified to a planar problem. This is achieved by evaluating if a set of integration points (see X in fig. \ref{fig:super_element_assignment}) is inside or outside the geometry and if they are inside or outside of the Super Element. Now, an assignment of all $N$ is searched which minimizes the sum of all $R$ for the selected Super Elements in their respective Precursor Mesh elements. The integration points can be tested for being inside the Super Elements once as the Super Elements are not problem dependent. We are using a grid of $5 \times 5 \times 5$ integration points as this is the smallest uniform grid which is able to differentiate all 256 Super Elements.

Computing $R$ as the sum of all deviating integration points leads to wavy mesh surfaces for plain geometries. We have found solutions to have a much higher quality when computing $R$ as $R = R_S^2 + R_P$ where $R_S$ is the sum of all integration points only occurring in the Super Element and $R_S$ are integration points only occurring in the geometry. By doing so cutting more material is favoured and the algorithm becomes more stable.

We solve the minimization problem of $R(N)$ with the \ac{ASP} solver Clingo to compute optimal solutions. 
We were able to solve the benchmark problem in Figure \ref{fig:retrenched_mesh} for up to 64 elements in the Precursor Mesh with proven optimality. 
For larger meshes no optimum was found in reasonable computation time and the non-optimal results were not satisfactory. 

In order to scale the overall method to real world instances we have developed a heuristic that computes good approximations for $R(N)$. 
This heuristic works by iterating over all elements of the Precursor Mesh, finding the most suitable Super Element for it. All elements vote whether they want adjacent nodes included or excluded from the mesh. The execution time of this algorithm is linear in element count. 

When comparing the \ac{ASP} based exact method to the heuristics, the two methods compute the same results on our set of sample instances. This indicates that our heuristic provides decent solutions for larger instances.

\section{Entity Mapping} \label{chapter:entity_mapping}

\begin{figure}
\begin{center}
\includegraphics[width=0.8\textwidth]{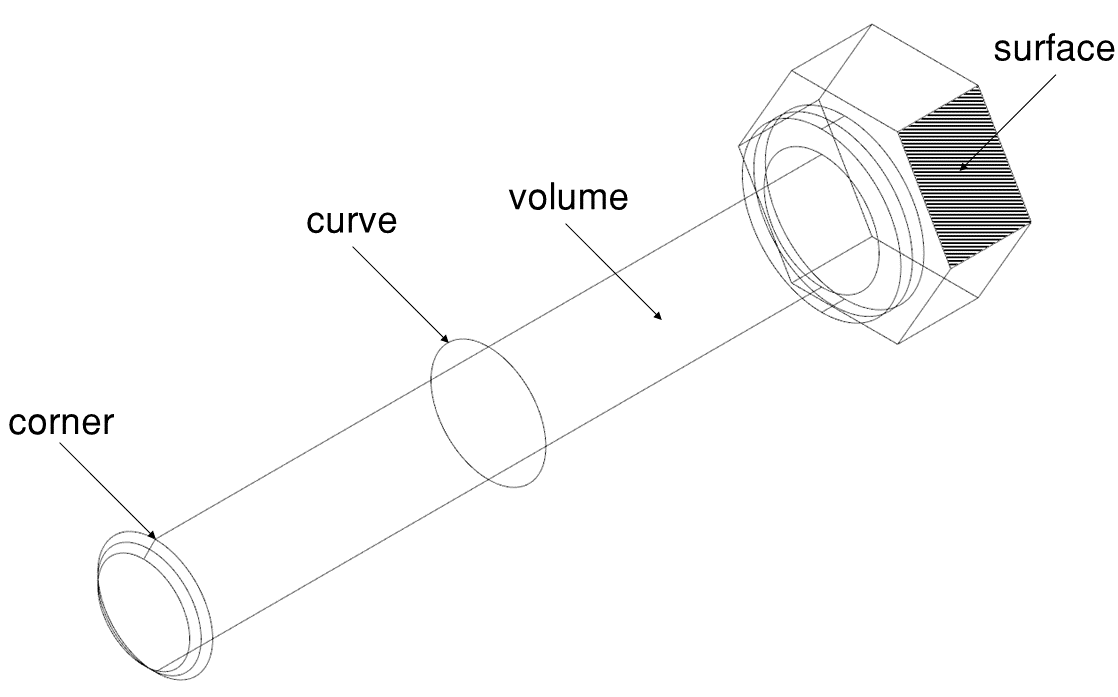}
\end{center}
\caption{Geometric entities}
\label{fig:geo_entyties}
\end{figure}

In the entity mapping phase of the algorithm, all surface nodes are assigned to geometric entities. This allows to improve geometric accuracy, elevate mesh order and insert more fine grained elements. Geometric entities are surfaces which enclose and define volumes. They are enclosed by curves which are terminated with corners as is shown in Figure \ref{fig:geo_entyties}. 

\begin{figure}
\centering
\def\svgwidth{0.75\textwidth}
\input{pictures/entity_mapping/entity_mapping}
\caption{Entity Mapping process}
\label{fig:entity_mapping}
\end{figure}
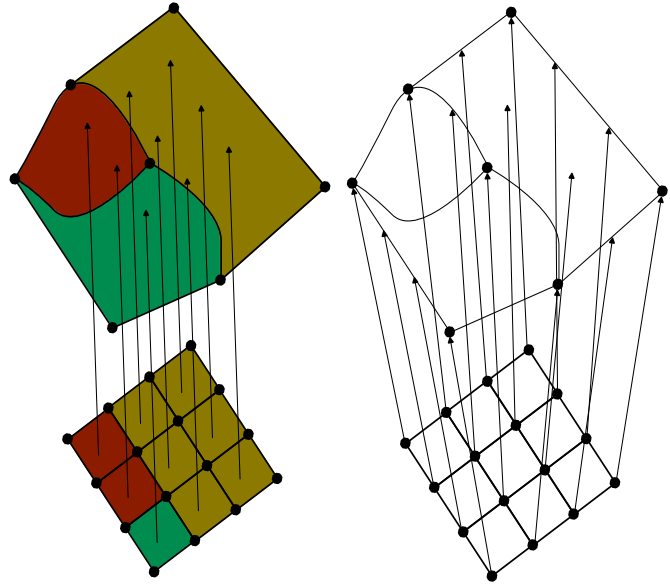

For each node of the surface mesh one geometric entity has to be assigned. We achieve this assignment by first assigning a surface to each element face of the surface mesh. The surface is determined by projecting a line from the element face center in normal direction. The surface assigned to the element face is the first one intersecting this line regardless of the line parameter sign of the intersection. In this process we obtain a colored mesh (see fig. \ref{fig:entity_mapping} left). We execute the intersection test with a fine triangular mesh of the target geometry because line intersection test with NURBS geometry are complex and unreliable The intersection point of a NURBS surface and a line is a nonlinear problem whereas the intersection problem of a line and a surface is a linear problem. The unreliability stems from the fact that there is no way to securely test that an arbitrary nonlinear function has no roots. The triangular mesh is generated with $\frac{1}{4}$ edge length of the target mesh. When all element faces which are adjacent to a node have the same color it is assigned to this surface. In case the adjacent element faces have two different colors the node is assigned to the curve splitting the two surfaces. If the two surfaces share multiple curves the closet curve to the node is chosen. When more then two colors occur, the node must be assigned to a corner. If there is a corner with the same colors in neighbouring surfaces it is selected. Otherwise the distance between node and potentially assigned corners is used as a tiebreaker.

In this process, any geometry smaller than the element size of the mesh is automatically de-featured.

A trivial implementation of the algorithm would require to project a test line from each face of the surface mesh to each triangle of the underlying tessellation. In the worst case, the count of surface mesh faces is the same order of magnitude as the volume element count and the total node count. In this case, the time complexity of the algorithm with respect to the total mesh nodes $n$ is in $\mathcal{O}(n^2)$.

This run-time can be improved to $\mathcal{O}(n \log n)$ by employing a more sophisticated ray casting method such as lazy sweep \cite{silvia_lazy_1997}. 

\section{Results on Examples} \label{chapter:results_on_exmples}

The method is tested with a subset of the MAMBO \cite{ledoux_mambo_2022} data set. The Precursor Mesh for the geometries is selected automatically by analysing the CAD tree structure of the model. The results are presented in Table \ref{tab:MAMBO}.
The resulting meshes and their quality reflect the underlying model structure of the geometry. B5 is modeled as a revolution whereas B14 is modeled as an extrusion. In the case of B11 and B14 the circular cross section is detected and a specialized surface mesher is used for the 2D mesh before sweeping. The elliptic cross section of B15 is not recognized resulting in a lower quality mesh.
The difference between B10 and our own example from fig. \ref{fig:retrenched_mesh} is caused by model structure as well. The MAMBO example is modeled as a single extrusion and our own example is modeled as an extruded cube and a subtracted cylinder.

In general our method show good performance on locally convex geometries. For cases such as B2, which contain sharp non convex features, the method fails to generate an adequate mesh. This issue is caused by the fact that the algorithm currently only uses convex Super Elements. The results can be improved by including non-convex Super Elements.

\begin{table*}
    \centering
    \begin{tabular}{cccc}
\includegraphics[width=0.24\textwidth]{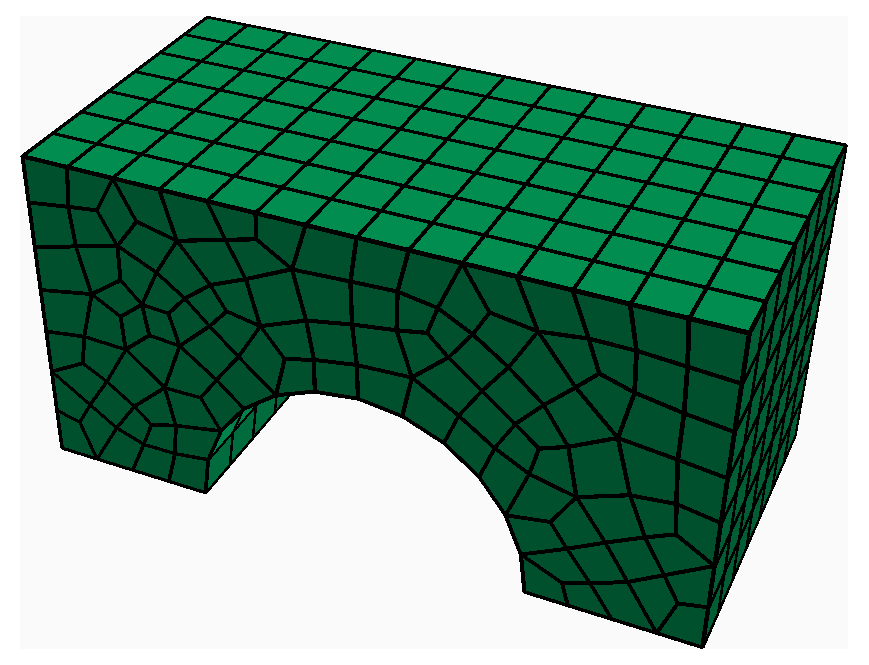} & 
\includegraphics[width=0.24\textwidth]{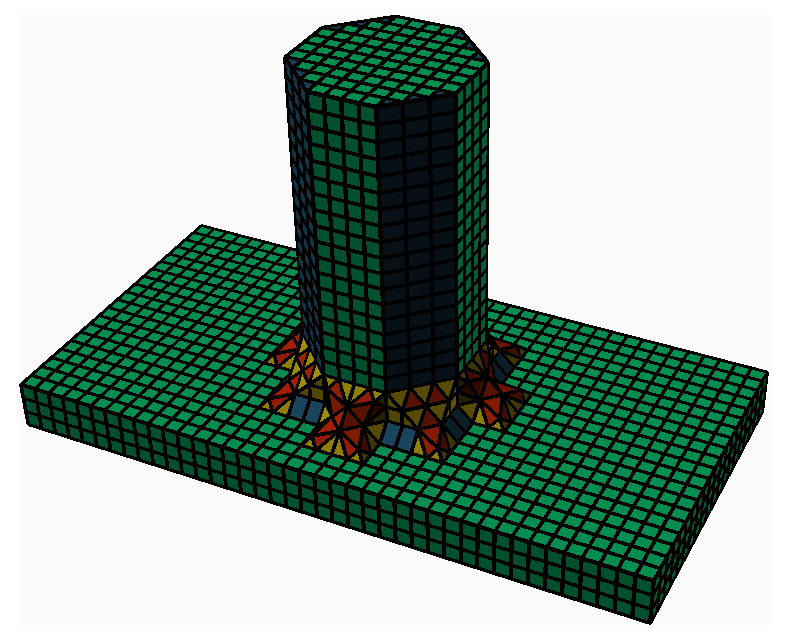} & 
\includegraphics[width=0.24\textwidth]{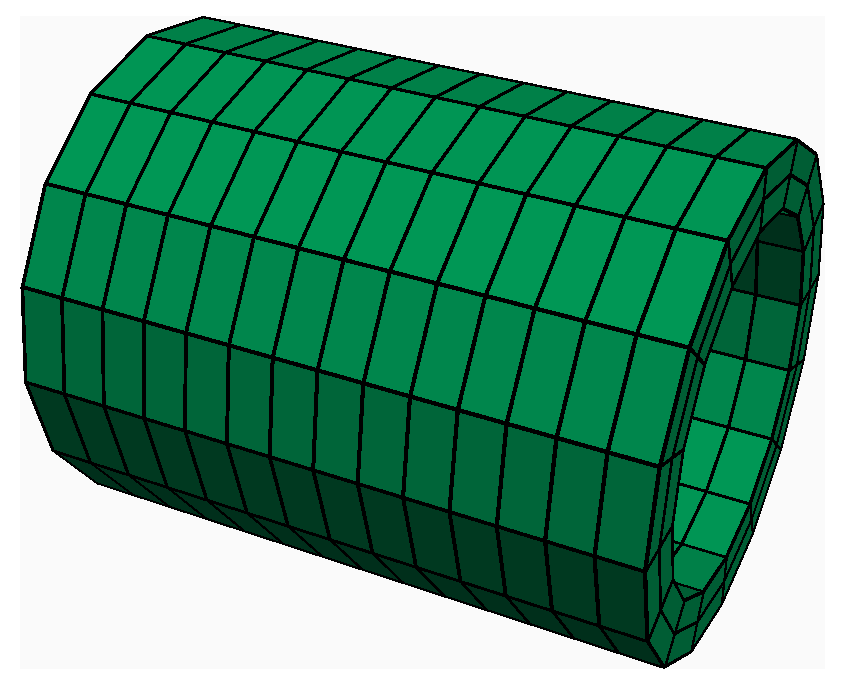} & 
\includegraphics[width=0.24\textwidth]{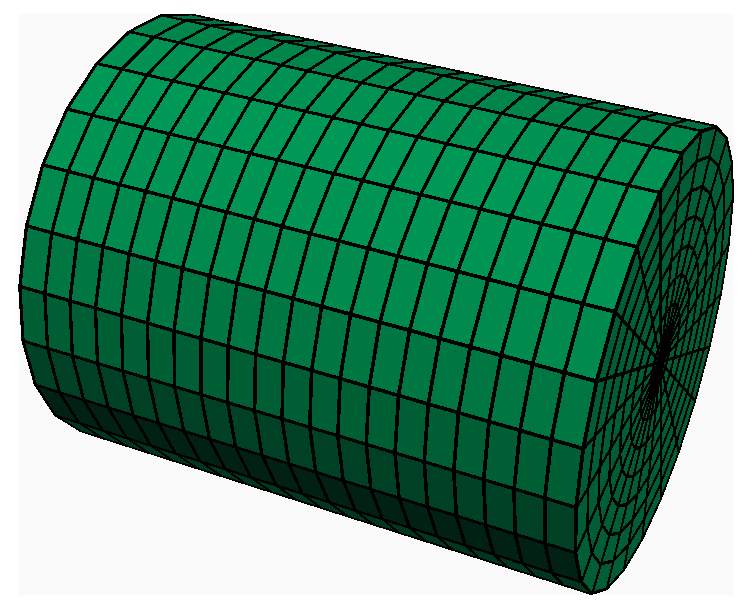} \\
B0 & B2 & B4 & B5 \\
\includegraphics[width=0.24\textwidth]{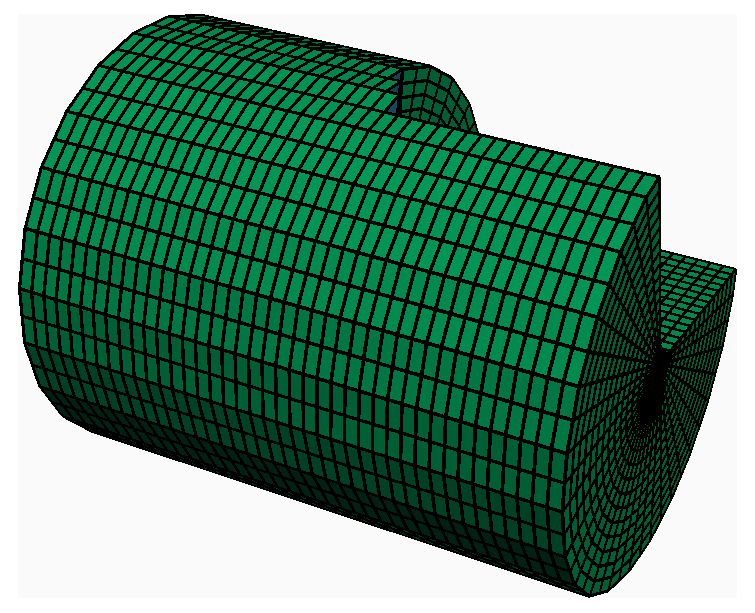} & 
\includegraphics[width=0.24\textwidth]{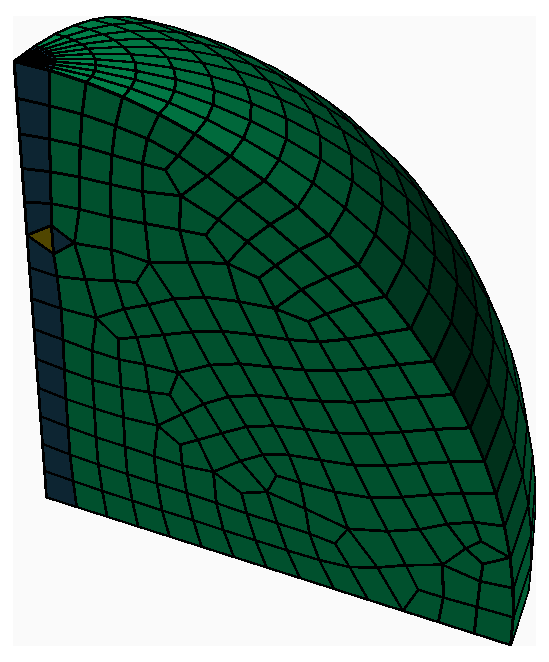} & 
\includegraphics[width=0.24\textwidth]{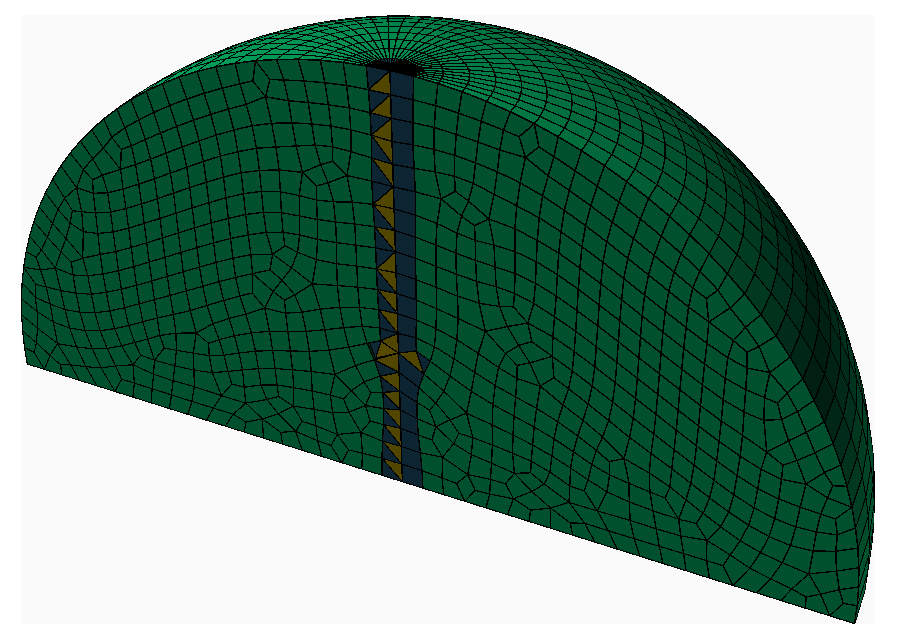} & 
\includegraphics[width=0.24\textwidth]{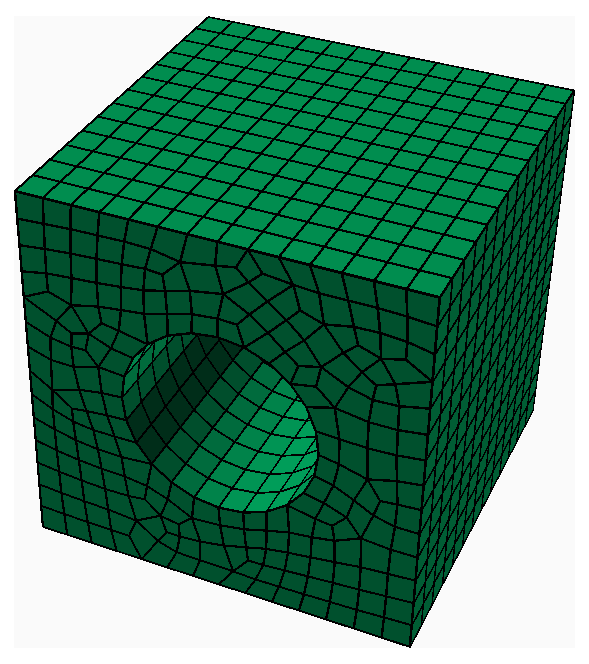} \\
B6 & B7 & B9 & B10 \\ 
\includegraphics[width=0.24\textwidth]{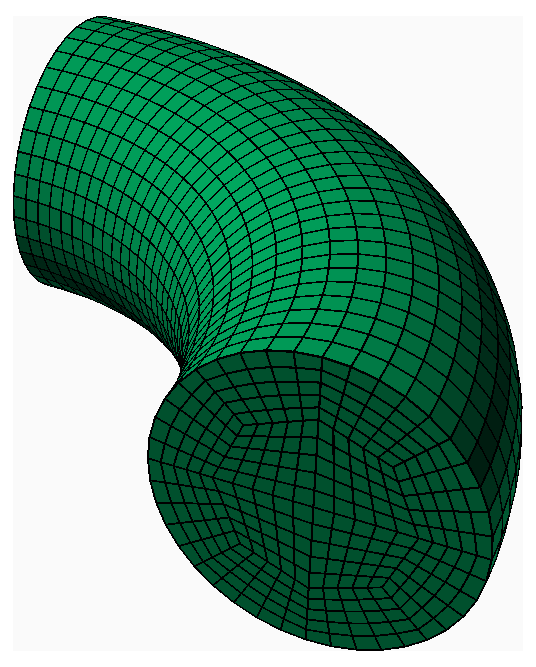} & 
\includegraphics[width=0.24\textwidth]{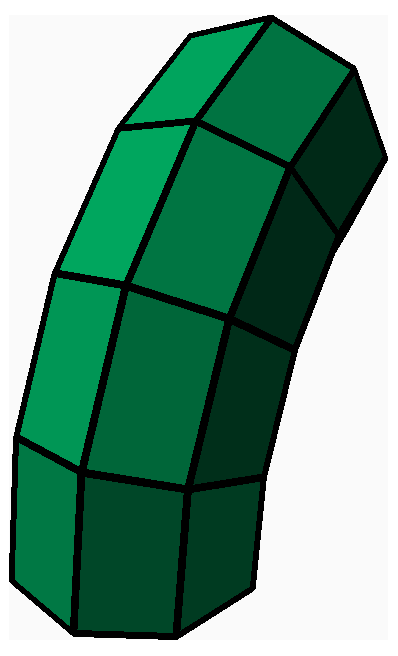} & 
\includegraphics[width=0.24\textwidth]{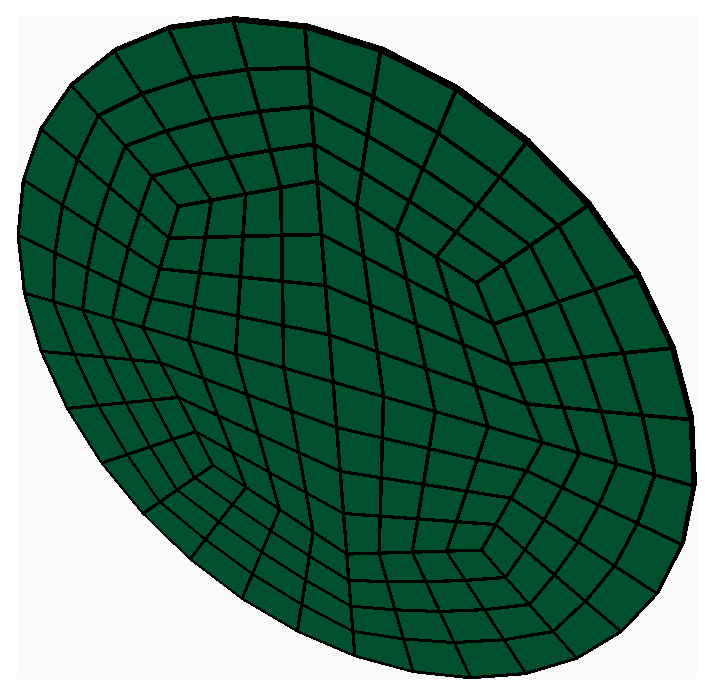} & 
\includegraphics[width=0.24\textwidth]{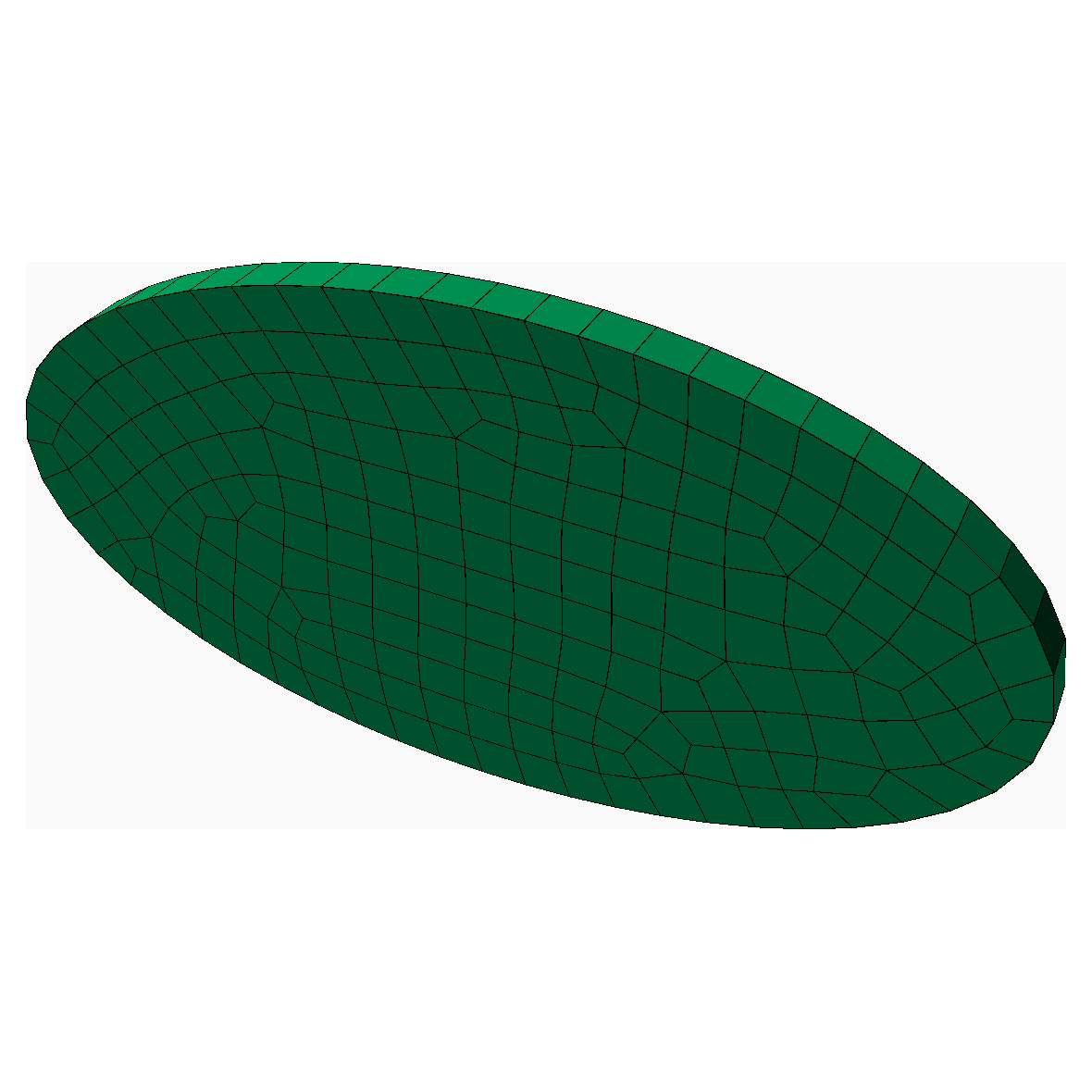} \\
B11 & B12 & B14 & B15 \\
\includegraphics[width=0.24\textwidth]{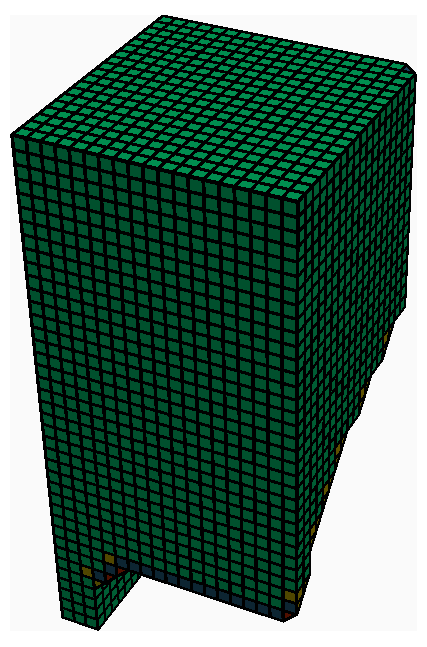} & 
\includegraphics[width=0.24\textwidth]{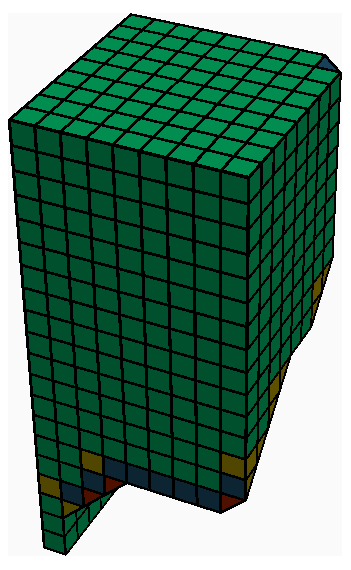} & 
\includegraphics[width=0.24\textwidth]{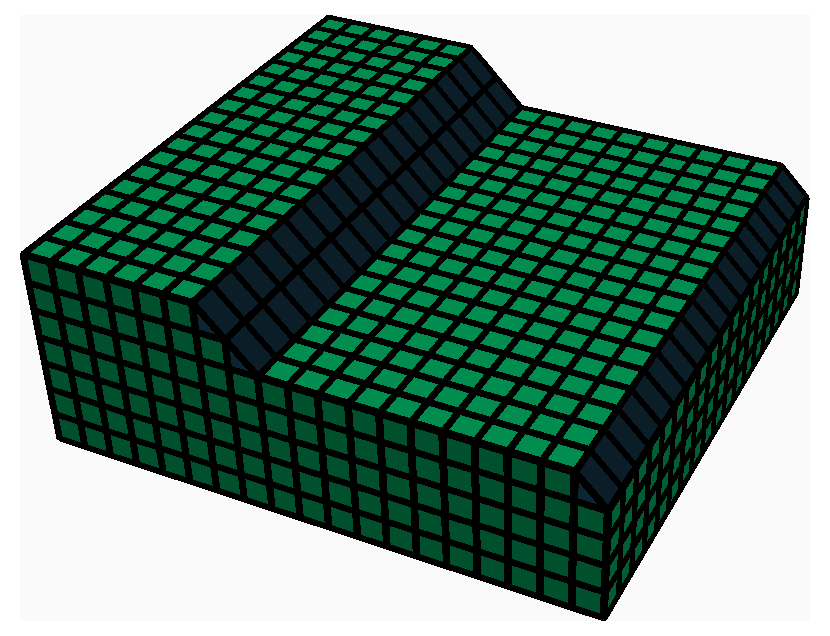} & 
\includegraphics[width=0.24\textwidth]{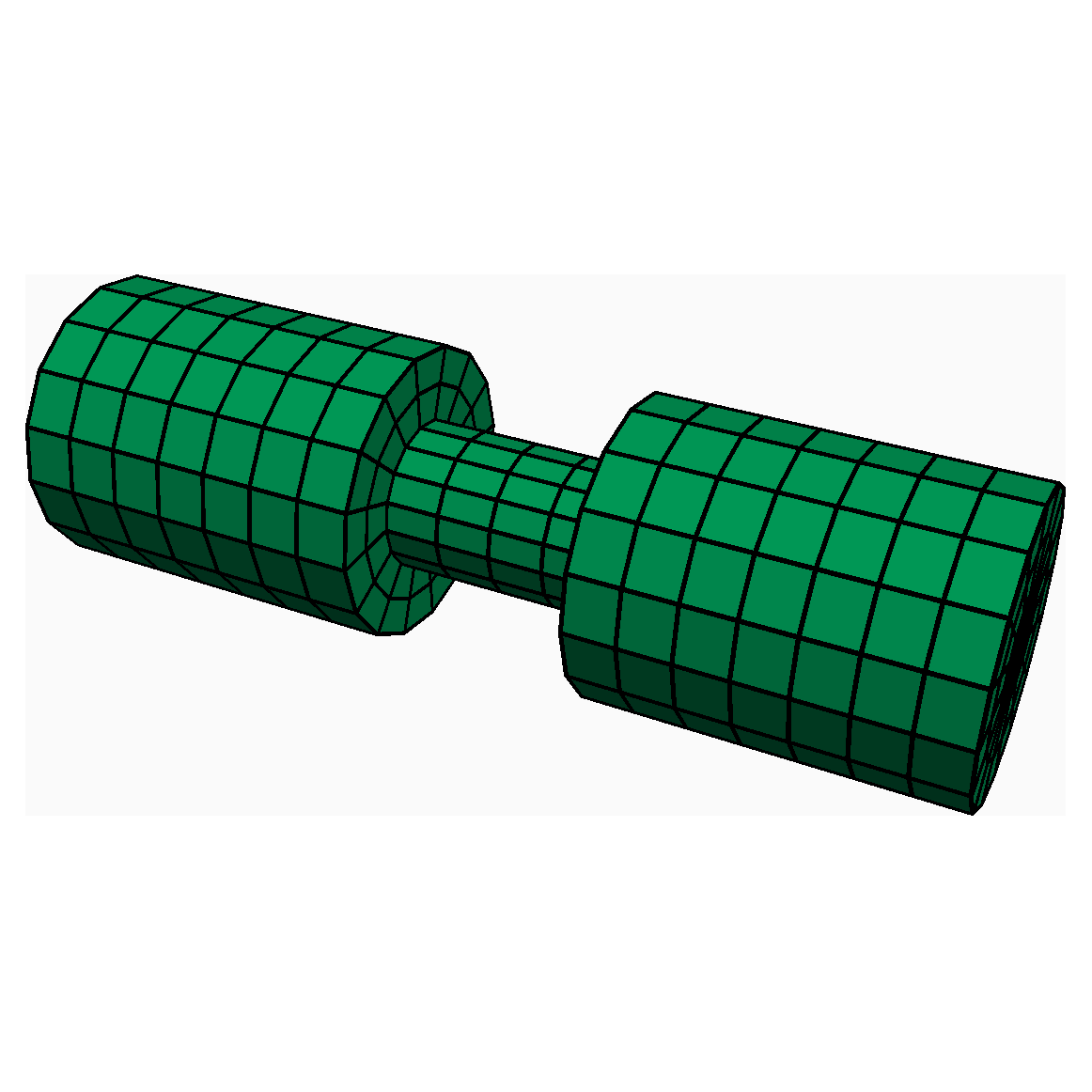} \\
B17 & B19 & B21 & B42 \\
\end{tabular}
    \caption{The hex-dominant meshes automatically computed by our method from basic instances of the MAMBO geometry benchmark.}
    \label{tab:MAMBO}
\end{table*}

\section{Conclusion and Future work}\label{chapter:futurework}

We have presented a novel approach for grid-based meshing using heuristic and combinatorial methods.
For two steps of our pipeline we applied Answer Set Programming to solve combinatorial sub-problem.
The set of schemes for subdivisions of the cube may find application in other hybrid meshing algorithms.   
Our success of using ASP in the context of meshing may inspire other combinatorial analysis of related problems.  

Our method advocates to use the underlying CAD model of the geometry. Especially in the context of FEM of mechanical engineering, we believe that this should receive more attention. Future algorithms could compare their solutions to ours using the same CAD tree from the MAMBO geometry benchmark. These CAD models and the ASP code we have used and all the computed Super Elements, are made available on Github: \url{https://github.com/HenrikJStromberg/combinatorial_meshing}

As next steps in our work, we see the following improvements. 
The current results of our mesh optimization seem to be inferior to those presented by \cite{owen_evaluation_2015}. We will combine their mesh optimization with our algorithm in the future.
We plan to evaluate our method on more complex instances available in the MAMBO set. Finally, we aim to improve the efficiency of our implementation to tackle problem sizes common in industrial applications.


\section{Acknowledgements} \label{chapter:acknowledgements}
We would like to thank Jeff Erickson and Jean Christoph Jung for discussing  our ideas with us and giving valuable feedback.

\bibliographystyle{siam}
\bibliography{p}
\end{document}

%% file: pictures/internal_nodes_hex/internal_nodes_hex.tex
\begingroup%
  \makeatletter%
  \providecommand\color[2][]{%
    \errmessage{(Inkscape) Color is used for the text in Inkscape, but the package 'color.sty' is not loaded}%
    \renewcommand\color[2][]{}%
  }%
  \providecommand\transparent[1]{%
    \errmessage{(Inkscape) Transparency is used (non-zero) for the text in Inkscape, but the package 'transparent.sty' is not loaded}%
    \renewcommand\transparent[1]{}%
  }%
  \providecommand\rotatebox[2]{#2}%
  \newcommand*\fsize{\dimexpr\f@size pt\relax}%
  \newcommand*\lineheight[1]{\fontsize{\fsize}{#1\fsize}\selectfont}%
  \ifx\svgwidth\undefined%
    \setlength{\unitlength}{297.00547586bp}%
    \ifx\svgscale\undefined%
      \relax%
    \else%
      \setlength{\unitlength}{\unitlength * \real{\svgscale}}%
    \fi%
  \else%
    \setlength{\unitlength}{\svgwidth}%
  \fi%
  \global\let\svgwidth\undefined%
  \global\let\svgscale\undefined%
  \makeatother%
  \begin{picture}(1,0.83321501)%
    \lineheight{1}%
    \setlength\tabcolsep{0pt}%
    \put(0,0){\includegraphics[width=\unitlength]{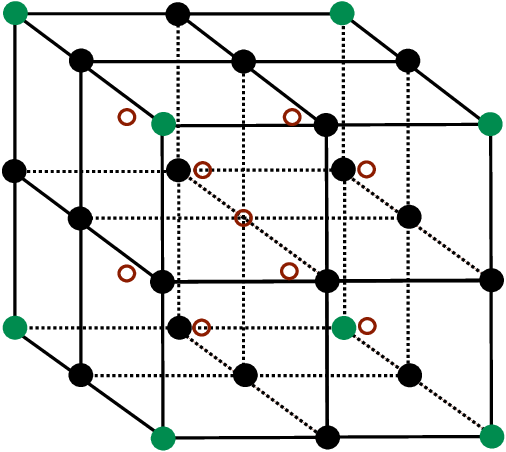}}%
  \end{picture}%
\endgroup%

%% file: pictures/mesh_graph/mesh_graph.tex
\begingroup%
  \makeatletter%
  \providecommand\color[2][]{%
    \errmessage{(Inkscape) Color is used for the text in Inkscape, but the package 'color.sty' is not loaded}%
    \renewcommand\color[2][]{}%
  }%
  \providecommand\transparent[1]{%
    \errmessage{(Inkscape) Transparency is used (non-zero) for the text in Inkscape, but the package 'transparent.sty' is not loaded}%
    \renewcommand\transparent[1]{}%
  }%
  \providecommand\rotatebox[2]{#2}%
  \newcommand*\fsize{\dimexpr\f@size pt\relax}%
  \newcommand*\lineheight[1]{\fontsize{\fsize}{#1\fsize}\selectfont}%
  \ifx\svgwidth\undefined%
    \setlength{\unitlength}{272.54487045bp}%
    \ifx\svgscale\undefined%
      \relax%
    \else%
      \setlength{\unitlength}{\unitlength * \real{\svgscale}}%
    \fi%
  \else%
    \setlength{\unitlength}{\svgwidth}%
  \fi%
  \global\let\svgwidth\undefined%
  \global\let\svgscale\undefined%
  \makeatother%
  \begin{picture}(1,1.32982053)%
    \lineheight{1}%
    \setlength\tabcolsep{0pt}%
    \put(0,0){\includegraphics[width=\unitlength]{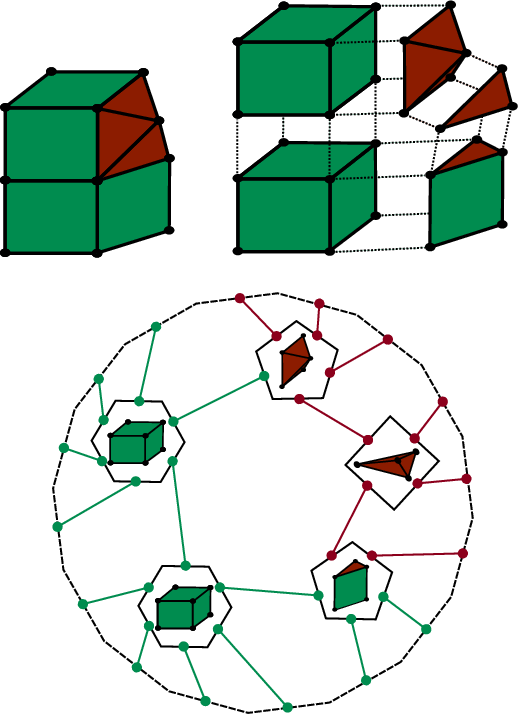}}%
  \end{picture}%
\endgroup%

%% file: pictures/super_eleme_assignement/super_eleme_assignement.tex
\begingroup%
  \makeatletter%
  \providecommand\color[2][]{%
    \errmessage{(Inkscape) Color is used for the text in Inkscape, but the package 'color.sty' is not loaded}%
    \renewcommand\color[2][]{}%
  }%
  \providecommand\transparent[1]{%
    \errmessage{(Inkscape) Transparency is used (non-zero) for the text in Inkscape, but the package 'transparent.sty' is not loaded}%
    \renewcommand\transparent[1]{}%
  }%
  \providecommand\rotatebox[2]{#2}%
  \newcommand*\fsize{\dimexpr\f@size pt\relax}%
  \newcommand*\lineheight[1]{\fontsize{\fsize}{#1\fsize}\selectfont}%
  \ifx\svgwidth\undefined%
    \setlength{\unitlength}{522.33476293bp}%
    \ifx\svgscale\undefined%
      \relax%
    \else%
      \setlength{\unitlength}{\unitlength * \real{\svgscale}}%
    \fi%
  \else%
    \setlength{\unitlength}{\svgwidth}%
  \fi%
  \global\let\svgwidth\undefined%
  \global\let\svgscale\undefined%
  \makeatother%
  \begin{picture}(1,0.54736741)%
    \lineheight{1}%
    \setlength\tabcolsep{0pt}%
    \put(0,0){\includegraphics[width=\unitlength]{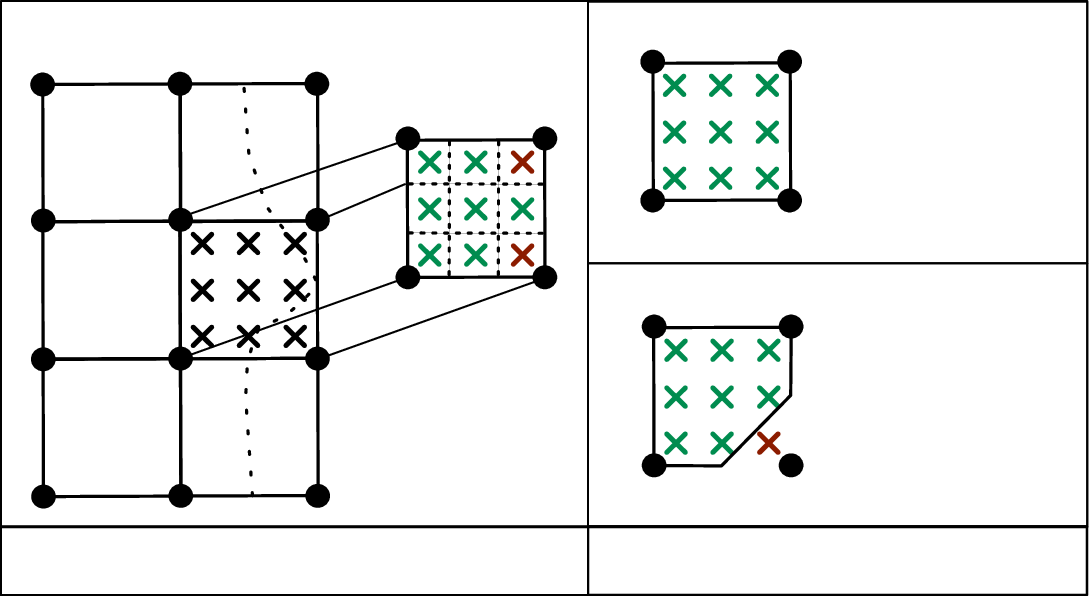}}%
    \put(0.7625842,0.4174569){\color[rgb]{0,0,0}\makebox(0,0)[lt]{\lineheight{1.25}\smash{\begin{tabular}[t]{l}$R = 4$\end{tabular}}}}%
    \put(0.7625842,0.17401111){\color[rgb]{0,0,0}\makebox(0,0)[lt]{\lineheight{1.25}\smash{\begin{tabular}[t]{l}$R = 1$\end{tabular}}}}%
    \put(0.13,0.02257685){\color[rgb]{0,0,0}\makebox(0,0)[lt]{\lineheight{1.25}\smash{\begin{tabular}[t]{l}Precursor Mesh\end{tabular}}}}%
    \put(0.55,0.02534147){\color[rgb]{0,0,0}\makebox(0,0)[lt]{\lineheight{1.25}\smash{\begin{tabular}[t]{l}Possibile Super Elements\end{tabular}}}}%
  \end{picture}%
\endgroup%

%% file: pictures/entity_mapping/entity_mapping.tex
\begingroup%
  \makeatletter%
  \providecommand\color[2][]{%
    \errmessage{(Inkscape) Color is used for the text in Inkscape, but the package 'color.sty' is not loaded}%
    \renewcommand\color[2][]{}%
  }%
  \providecommand\transparent[1]{%
    \errmessage{(Inkscape) Transparency is used (non-zero) for the text in Inkscape, but the package 'transparent.sty' is not loaded}%
    \renewcommand\transparent[1]{}%
  }%
  \providecommand\rotatebox[2]{#2}%
  \newcommand*\fsize{\dimexpr\f@size pt\relax}%
  \newcommand*\lineheight[1]{\fontsize{\fsize}{#1\fsize}\selectfont}%
  \ifx\svgwidth\undefined%
    \setlength{\unitlength}{569.18072558bp}%
    \ifx\svgscale\undefined%
      \relax%
    \else%
      \setlength{\unitlength}{\unitlength * \real{\svgscale}}%
    \fi%
  \else%
    \setlength{\unitlength}{\svgwidth}%
  \fi%
  \global\let\svgwidth\undefined%
  \global\let\svgscale\undefined%
  \makeatother%
  \begin{picture}(1,0.8800246)%
    \lineheight{1}%
    \setlength\tabcolsep{0pt}%
    \put(0,0){\includegraphics[width=\unitlength]{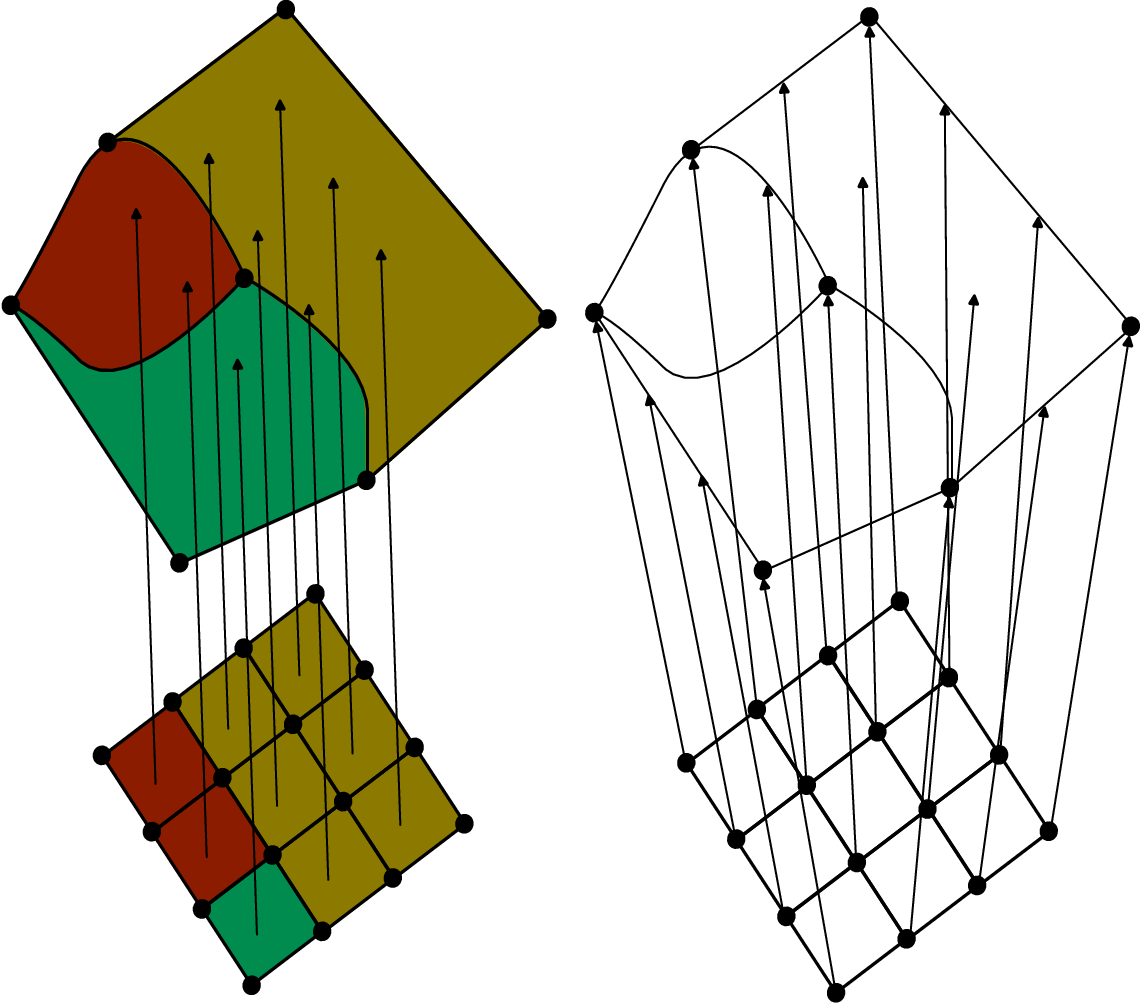}}%
  \end{picture}%
\endgroup%